\documentclass[aps, prd, twocolumn, superscriptaddress, nofootinbib]{revtex4-1}

\usepackage{latexsym}
\usepackage{amsmath}
\usepackage{amssymb}
\usepackage{amsfonts}
\usepackage{bm}

\usepackage{color}

\usepackage{supertabular} 
\usepackage{placeins}
\usepackage{epsfig}
\usepackage{graphicx}

\begin{document}

%\preprint{}

\title{Doubly Charmed Pentaquarks}


\author{Gang Yang}
\email[]{ygz0788a@sina.com}
%\thanks{}
\affiliation{Department of Physics and State Key Laboratory of Low-Dimensional Quantum Physics, \\ Tsinghua University, Beijing 100084, China}

\author{Jialun Ping}
\email[]{jlping@njnu.edu.cn}
\affiliation{Department of Physics and Jiangsu Key Laboratory for Numerical Simulation of Large Scale Complex Systems, \\ Nanjing Normal University, Nanjing 210023, P. R. China}

\author{Jorge Segovia}
\email[]{jsegovia@upo.es}
\affiliation{Departamento de Sistemas F\'isicos, Qu\'imicos y Naturales, \\ Universidad Pablo de Olavide, E-41013 Sevilla, Spain}
%\affiliation{Dpto. Sistemas F\'isicos, Qu\'imicos y Naturales, Univ. Pablo de Olavide, 41013 Sevilla, Spain}
%\affiliation{Institute for Nonperturbative Physics, Nanjing University, Nanjing, Jiangsu 210093, China}


\date{\today}

\begin{abstract}
The LHCb Collaboration, using its full data set from runs~$1$ and~$2$, announced in $2019$ a surprising update of the hidden-charm pentaquark states $P_c(4380)^+$ and $P_c(4450)^+$, observed in 2015. A new state, $P_c(4312)^+$, was clearly seen at lower energies; furthermore, the original $P_c(4450)$ resonance was resolved into two individual states, named the $P_c(4440)^+$ and the $P_c(4457)^+$.
Motivated by the fact that these new hidden-charm pentaquark states were successfully predicted by our chiral quark model, we extend herein such study to the doubly charmed sector. The analyzed total spin and parity quantum numbers are $J^P=\frac12^-$, $\frac32^-$ and $\frac52^-$, in the $I=\frac12$ and $\frac32$ isospin channels.
We find several possible narrow baryon-meson resonances (theoretical masses in parenthesis): $IJ^P = \frac12 \frac12^-$ $\Sigma_c D(4356)$, $\frac12 \frac32^-$ $\Sigma^*_c D(4449)$, $\frac32 \frac12^-$ $\Sigma_c D(4431)$, $\frac32 \frac12^-$ $\Sigma_c D(4446)$, $\frac32 \frac32^-$ $\Sigma_c D^*(4514)$ and $\frac32 \frac52^-$ $\Xi^*_{cc} \rho(4461)$ whose widths are $4.8$, $8.0$, $2.6$, $2.2$, $4.0$ and $3.0\,\text{MeV}$, respectively. Moreover, one shallow bound-state is found, too, with quantum numbers $IJ^P = \frac12 \frac32^-$ $\Xi^*_{cc} \pi(3757)$. These doubly charmed pentaquark states are expected to be identified in future experiments.
\end{abstract}

\pacs{
12.38.-t \and % Quantum Chromodynamics
12.39.-x \and % Potential Models
14.20.-c \and % Properties of Baryons
14.20.Pt      % Exotic Baryons
}
\keywords{
Quantum Chromodynamics \and
Quark models           \and
Properties of Baryons  \and
Exotic Baryons
}

\maketitle

%%%%%%%%%%%%%%%%%%%%%%%%%%%%%%%%%%%%%%%%%%%%%%%%%%%%%%%%%%%%%%%%%%%%%%%%%%%%%%%%

\section{Introduction}

During the past $15$ years, more than two dozens of nontraditional charmonium- and bottomonium-like states, the so-called XYZ mesons, have been observed at B-factories (BaBar, Belle and CLEO), $\tau$-charm facilities (CLEO-c and BESIII) and also proton-(anti)proton colliders (CDF, D0, LHCb, ATLAS and CMS). Among all of them, one can highlight the new three hidden-charm pentaquark candidates observed in 2019 by the LHCb Collaboration~\cite{lhcb:2019pc} in the $J/\psi p$ invariant mass spectrum of $\Lambda^{0}_{b} \rightarrow J/\psi K^{-}p$ decays, they are signed as $P_c(4312)^+$, $P_c(4440)^+$ and $P_c(4457)^+$, respectively. The story of hidden-charm pentaquark states can actually be dated back to $2015$, when two exotic signals: $P_c(4380)^+$ and $P_c(4450)^+$, were announced by the same collaboration~\cite{Aaij:2015tga}. Two striking features characterized these states: they appear quite close to baryon-meson thresholds and all are very narrow; this is believed to be an invaluable information towards discriminating between different explanations on how the quarks are arranged within the pentaquarks. 

There is an intensive theoretical activity on explaining the dynamical mechanism that produces the three newly observed hidden-charm pentaquarks, $P_c(4312)^+$, $P_c(4440)^+$ and $P_c(4457)^+$. A common one is the baryon-meson molecular picture, {\it i.e.} $\Sigma_c \bar{D}^{(*)}$ states described within different kind of formalisms such as effective field theories~\cite{MZL190311560, JH190311872}, heavy quark spin symmetry approach~\cite{YS190400587, CWX190401296}, phenomenological potential models~\cite{ZHG190400851, HH190400221, HM1904.09756, RZ190410285, MIE190411616, XZW190409891, FLW190503636}, heavy hadron chiral perturbation theory~\cite{LM190504113}, and QCD sum rules~\cite{ZGW190502892, JRZ190410711}. The $P_c(4312)^+$ and $P_c(4457)^+$ signals have been studied independently in Refs.~\cite{CFR190410021} and~\cite{FKG190311503} using the S-matrix method but in the later case through isospin-violating decay channels. Moreover, the decay properties of the three $P_c^+$ states have been computed in Ref.~\cite{CJX190400872}, and their photo-production has been interestingly discussed in Refs.~\cite{XC190406015, XYW190411706}. 

It is important to highlight here that, before the LHCb's announcement of the three new hidden-charm pentaquark states, their existence were predicted by some of the present authors in Ref.~\cite{Yang:2015bmv} (see Tables III and IV). The $P^+_c(4312)$, $P^+_c(4440)$ and $P^+_c(4457)$ were described as baryon-meson molecular states of the form $J^P=\frac12^-$ $\Sigma_c\bar{D}$, $\frac12^-$ $\Sigma_c\bar{D}^*$ and $\frac32^-$ $\Sigma_c\bar{D}^*$, respectively; belonging all of them to the isospin $I=\frac12$ sector. Moreover, these results are supported by other theoretical studies such as the ones reported in Refs.~\cite{MZL190311560, JH190311872, CWX190401296, CJX190400872}.

Apart from the hidden-charm pentaquark states, there are also other pentaquark configurations triggering theoretical interest. One heavy antiquark pentaquarks, $\bar{Q}qqqq$, are analyzed within a constituent quark model and no bound-state is found~\cite{JMR190103578}. Doubly heavy pentaquarks are systematically studied in a phenomenological potential model with the conclusion that either stable states or narrow resonances are possible~\cite{QSZ180104557, FG190304430}. In Ref.~\cite{LM181107320}, light pseudoscalar meson and doubly charmed baryon scattering lengths are calculated by means of the heavy baryon chiral perturbation theory. Possible triply charmed molecular pentaquarks such as $\Xi_{cc}D_1(\bar{D}_1)$ and $\Xi_{cc}D_2^*(\bar{D}_2^*)$ are proposed using a one-boson-exchange model in Ref.~\cite{FLW190101542}; and the mass splittings for the $S$-wave triply heavy pentaquark states are systematically calculated~\cite{HTA190507858}.  Meanwhile, some interesting reviews discussing the pentaquark issue but also collecting information about, {\it e.g.}, tetraquark states can be found in Refs.~\cite{JV190209799, YRL190311976}; moreover, potential prospects on the production of multiquark systems containing heavy quarks with the ALICE experiment at LHC are discussed in Ref.~\cite{RV190406180}.

Within a chiral quark model formalism\footnote{This approach has been successfully applied to the charmonium, bottomonium and heavy baryon sectors, studying their spectra~\cite{Segovia:2008zz, Segovia:2016xqb, Yang:2019lsg}, their electromagnetic, weak and strong decays and reactions~\cite{Segovia:2011zza, Segovia:2013kg, Segovia:2014mca}, and their coupling with meson-meson thresholds~\cite{Ortega:2009hj, Ortega:2016hde, Ortega:2017qmg}.}, we systematically study herein the possibility of having either bound or resonance states in the doubly charmed pentaquark sector with quantum numbers $J^P=\frac12^-$, $\frac32^-$ and $\frac52^-$, and in the $I=\frac12$ and $\frac32$ isospin sectors. This 5-body bound state problem is solved by means of the Gau\ss ian expansion method (GEM)~\cite{Hiyama:2003cu}, which has been demonstrated to be as accurate as a Faddeev calculation (see Figs.~15 and~16 of Ref.~\cite{Hiyama:2003cu}). Note, too, that the same approach has been applied in previous studies of $P_c$~\cite{Yang:2015bmv} and $P_b$ states~\cite{Yang:2018oqd}. 

\begin{figure}[ht]
\epsfxsize=3.4in \epsfbox{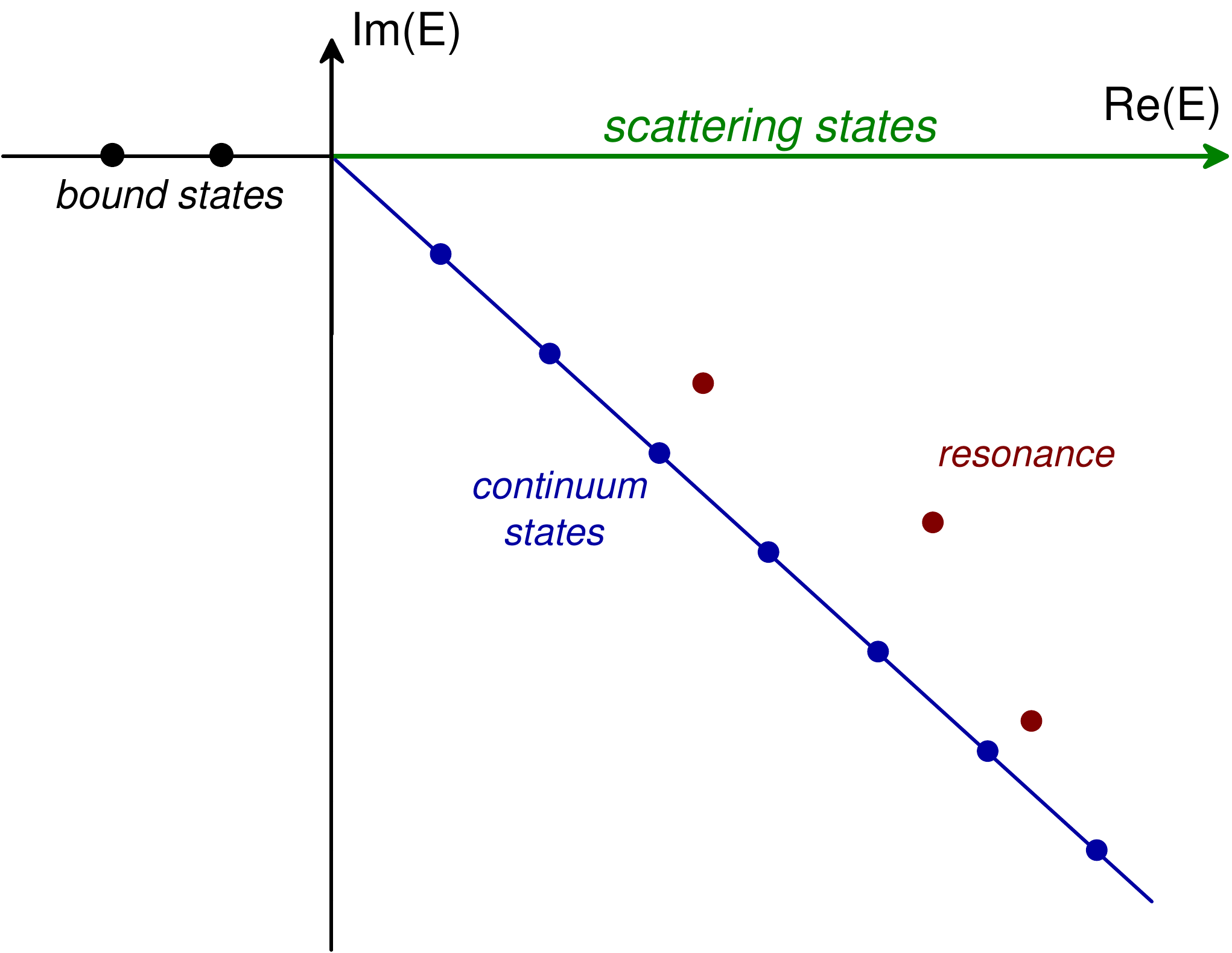}
\caption{\label{CSM1} Schematic complex energy distribution in the single-channel two-body system.}
\end{figure}

In this work, a powerful technique named complex scaling method (CSM) is employed in order to disentangle bound, resonance, and continuum (scattering) states within the same calculation. As illustration, Fig.~\ref{CSM1} shows a schematic distribution of the complex energy 2-body states obtained by the CSM, according to Ref.~\cite{TMPPNP7912014}. As one can see, the resonance states can be computed as an equivalent bound-state problem without resorting to the Lippmann-Schwinger equation formalism. Up to our knowledge, this is the first time that the CSM is applied to study pentaquark systems. During the past decades, CSM has been extensively applied to nuclear physics problems~\cite{SAPTP11612006, TMPPNP7912014}, and recently to the study of charmed di-baryon resonances~\cite{MO190400586} and doubly-heavy tetraquarks~\cite{Yang:2019itm}.

The structure of the present work is organized in the following way. In Sec.~\ref{sec:model} the ChQM, pentaquark wave-functions, GEM and CSM are briefly presented and discussed. Section~\ref{sec:results} is devoted to the analysis and discussion of our theoretical results. We summarize and give some prospects in Sec.~\ref{sec:summary}.

%%%%%%%%%%%%%%%%%%%%%%%%%%%%%%%%%%%%%%%%%%%%%%%%%%%%%%%%%%%%%%%%%%%%%%%%%%%%%%%%

\section{Theoretical framework}
\label{sec:model}

Lattice-QCD (LQCD) has made in the last decade or so an impressive progress on understanding multiquark systems~\cite{Alexandrou:2001ip, Okiharu:2004wy} and meson-meson, meson-baryon and baryon-baryon interactions~\cite{Prelovsek:2014swa, Lang:2014yfa, Briceno:2017max}; however, QCD-inspired quark models are still the main tool to shed some light on the nature of the multiquark candidates observed by experimentalists.

The general form of our five-body Hamiltonian, within the CSM approach, is given by
\begin{equation}
H(\theta) = \sum_{i=1}^{5}\left( m_i+\frac{\vec{p\,}^2_i}{2m_i}\right) - T_{\text{CM}} + \sum_{j>i=1}^{5} V(\vec{r}_{ij} e^{i\theta}) \,,
\label{eq:Hamiltonian}
\end{equation}
where each quark is considered nonrelativistic, $T_{\text{CM}}$ is the center-of-mass kinetic energy and the two-body potential
\begin{equation}\label{CQMV}
V(\vec{r}_{ij} e^{i\theta}) = V_{\text{CON}}(\vec{r}_{ij} e^{i\theta}) + V_{\text{OGE}}(\vec{r}_{ij} e^{i\theta}) + V_{\chi}(\vec{r}_{ij} e^{i\theta}) \,,
\end{equation}
includes color-confining, one-gluon-exchange and Goldstone-boson-exchange interactions. Herein, the coordinates of relative motions between quarks are transformed with a complex rotation, $\vec{r} \rightarrow \vec{r} e^{i\theta}$. Therefore, in the framework of complex range, the five-body systems are solved in a complex scaled Schr\"{o}dinger equation:
\begin{equation}\label{CSMSE}
\left[ H(\theta)-E(\theta) \right] \Psi_{JM}(\theta)=0 \,.
\end{equation}

According to the so-called ABC theorem~\cite{JA22269, EB22280}, there are three types of complex eigenenergies of Eq.~(\ref{CSMSE}), as shown in Fig.~\ref{CSM1}:
\begin{itemize}
\item Bound states below threshold are always located on the energy's negative real axis.
\item Discretized continuum states are aligned along the cut line with a rotated angle of 2$\theta$, related to the real axis.
\item Resonance states are fixed poles under the complex scaling transformation, and they are located above the continuum cut line. The resonance's width is given by $\Gamma=-2\,\text{Im}(E)$.
\end{itemize}

Coming back to the quark--(anti-)quark interacting potentials shown in Eq.~(\ref{CQMV}). Color confinement should be encoded in the non-Abelian character of QCD. LQCD studies have demonstrated that multi-gluon exchanges produce an attractive linearly rising potential proportional to the distance between infinite-heavy quarks~\cite{Bali:2005fu}. However, the spontaneous creation of light-quark pairs from the QCD vacuum may give rise at the same scale to a breakup of the color flux-tube~\cite{Bali:2005fu}. We have tried to mimic these two phenomenological observations by the following expression, in complex scaling:
\begin{equation}
V_{\text{CON}}(\vec{r}_{ij} e^{i\theta}\,)=\left[-a_{c}(1-e^{-\mu_{c}r_{ij} e^{i\theta}})+\Delta \right] 
(\vec{\lambda}_{i}^{c}\cdot\vec{\lambda}_{j}^{c}) \,,
\label{eq:conf}
\end{equation}
where $a_{c}$ and $\mu_{c}$ are model parameters, and the SU(3) color Gell-Mann matrices are denoted as $\lambda^c$. One can see in Eq.~\eqref{eq:conf} that the potential is linear at short inter-quark distances with an effective confinement strength $\sigma = -a_{c} \, \mu_{c} \, (\vec{\lambda}^{c}_{i}\cdot \vec{\lambda}^{c}_{j})$, while it becomes constant at large distances. 

The one-gluon-exchange potential contains central, tensor and spin-orbit contributions. We consider only the central term but also with a complex transformation, $\vec{r} \rightarrow \vec{r} e^{i\theta}$:
\begin{align}
V_{\text{OGE}}(\vec{r}_{ij} e^{i\theta}) &= \frac{1}{4} \alpha_{s} (\vec{\lambda}_{i}^{c}\cdot
\vec{\lambda}_{j}^{c}) \Bigg[\frac{1}{r_{ij} e^{i\theta}} \nonumber \\ 
&
- \frac{1}{6m_{i}m_{j}} (\vec{\sigma}_{i}\cdot\vec{\sigma}_{j}) 
\frac{e^{-r_{ij} e^{i\theta}/r_{0}(\mu)}}{r_{ij} e^{i\theta} r_{0}^{2}(\mu)} \Bigg] \,,
\end{align}
where $m_{i}$ is the quark mass and the Pauli matrices are denoted by $\vec{\sigma}$. The contact term of the central potential has been regularized as
\begin{equation}
\delta(\vec{r}_{ij} e^{i\theta})\sim\frac{1}{4\pi r_{0}^{2}}\frac{e^{-r_{ij} e^{i\theta}/r_{0}}}{r_{ij} e^{i\theta}} \,,
\end{equation}
with $r_{0}(\mu_{ij})=\hat{r}_{0}/\mu_{ij}$ a regulator that depends on $\mu_{ij}$, the reduced mass of the quark--(anti-)quark pair.

The wide energy range needed to provide a consistent description of mesons and baryons from light to heavy quark sectors requires an effective scale-dependent strong coupling constant. We use the frozen coupling constant of, for instance, Ref.~\cite{Segovia:2013wma}
\begin{equation}
\alpha_{s}(\mu_{ij})=\frac{\alpha_{0}}{\ln\left(\frac{\mu_{ij}^{2}+\mu_{0}^{2}}{\Lambda_{0}^{2}} \right)} \,,
\end{equation}
in which $\alpha_{0}$, $\mu_{0}$ and $\Lambda_{0}$ are parameters of the model.

The central terms of the chiral quark--(anti-)quark interaction with CSM can be written as
\begin{align}
&
V_{\pi}\left( \vec{r}_{ij} e^{i\theta} \right) = \frac{g_{ch}^{2}}{4\pi}
\frac{m_{\pi}^2}{12m_{i}m_{j}} \frac{\Lambda_{\pi}^{2}}{\Lambda_{\pi}^{2}-m_{\pi}
^{2}}m_{\pi} \Bigg[ Y(m_{\pi}r_{ij} e^{i\theta}) \nonumber \\
&
\hspace*{1.20cm} - \frac{\Lambda_{\pi}^{3}}{m_{\pi}^{3}}
Y(\Lambda_{\pi}r_{ij} e^{i\theta}) \bigg] (\vec{\sigma}_{i}\cdot\vec{\sigma}_{j})\sum_{a=1}^{3}(\lambda_{i}^{a}
\cdot\lambda_{j}^{a}) \,, \\
& 
V_{\sigma}\left( \vec{r}_{ij} e^{i\theta} \right) = - \frac{g_{ch}^{2}}{4\pi}
\frac{\Lambda_{\sigma}^{2}}{\Lambda_{\sigma}^{2}-m_{\sigma}^{2}}m_{\sigma} \Bigg[
Y(m_{\sigma}r_{ij} e^{i\theta}) \nonumber \\
&
\hspace*{1.20cm} - \frac{\Lambda_{\sigma}}{m_{\sigma}}Y(\Lambda_{\sigma}r_{ij} e^{i\theta})
\Bigg] \,, \\
& 
V_{K}\left( \vec{r}_{ij} e^{i\theta} \right)= \frac{g_{ch}^{2}}{4\pi}
\frac{m_{K}^2}{12m_{i}m_{j}} \frac{\Lambda_{K}^{2}}{\Lambda_{K}^{2}-m_{K}^{2}}m_{
K} \Bigg[ Y(m_{K}r_{ij} e^{i\theta}) \nonumber \\
&
\hspace*{1.20cm} -\frac{\Lambda_{K}^{3}}{m_{K}^{3}}Y(\Lambda_{K}r_{ij} e^{i\theta})
\Bigg] (\vec{\sigma}_{i}\cdot\vec{\sigma}_{j})\sum_{a=4}^{7}(\lambda_{i}^{a}
\cdot\lambda_{j}^{a}) \,, \\
& 
V_{\eta}\left( \vec{r}_{ij} e^{i\theta} \right) = \frac{g_{ch}^{2}}{4\pi}
\frac{m_{\eta}^2}{12m_{i}m_{j}} \frac{\Lambda_{\eta}^{2}}{\Lambda_{\eta}^{2}-m_{
\eta}^{2}}m_{\eta} \Bigg[ Y(m_{\eta}r_{ij} e^{i\theta}) \nonumber \\
&
\hspace*{1.20cm} -\frac{\Lambda_{\eta}^{3}}{m_{\eta}^{3}
}Y(\Lambda_{\eta}r_{ij} e^{i\theta}) \Bigg] (\vec{\sigma}_{i}\cdot\vec{\sigma}_{j})
\Big[\cos\theta_{p} \left(\lambda_{i}^{8}\cdot\lambda_{j}^{8}
\right) \nonumber \\
&
\hspace*{1.20cm} -\sin\theta_{p} \Big] \,,
\end{align}
where $Y(x)$ is the standard Yukawa function defined by $Y(x)=e^{-x}/x$. We consider the physical $\eta$ meson instead of the octet one and so we introduce the angle $\theta_p$. The $\lambda^{a}$ are the SU(3) flavor Gell-Mann matrices. Taken from their experimental values, $m_{\pi}$, $m_{K}$ and $m_{\eta}$ are the masses of the SU(3) Goldstone bosons. The value of $m_{\sigma}$ is determined through the PCAC relation $m_{\sigma}^{2}\simeq m_{\pi}^{2}+4m_{u,d}^{2}$~\cite{Scadron:1982eg}. Finally, the chiral coupling constant, $g_{ch}$, is determined from the $\pi NN$ coupling constant through
\begin{equation}
\frac{g_{ch}^{2}}{4\pi}=\frac{9}{25}\frac{g_{\pi NN}^{2}}{4\pi} \frac{m_{u,d}^{2}}{m_{N}^2} \,,
\end{equation}
which assumes that flavor SU(3) is an exact symmetry, only broken by the different mass of the strange quark.

As it is well known, the quark model parameters are crucial. In our case, the model parameters have been taken from, {\it e.g.}, Ref.~\cite{Yang:2015bmv} and, for completeness, they are listed in Table~\ref{model}. Note that the same set of model parameters was used in Refs.~\cite{Yang:2015bmv} and~\cite{Yang:2018oqd} to study, respectively, possible hidden-charm and -bottom pentaquark bound- and resonance-states.

\begin{table}[!t]
\caption{\label{model} Quark model parameters.}
\begin{ruledtabular}
\begin{tabular}{cccc}
Quark masses     & $m_u=m_d$ (MeV) &  313 \\
                 & $m_b$ (MeV)     & 5100 \\[2ex]
Goldstone bosons & $\Lambda_\pi=\Lambda_\sigma~$ (fm$^{-1}$) &   4.20 \\
                 & $\Lambda_\eta$ (fm$^{-1}$)     &   5.20 \\
                 & $g^2_{ch}/(4\pi)$                         &   0.54 \\
                 & $\theta_P(^\circ)$                        & -15 \\[2ex]
Confinement      & $a_c$ (MeV)         & 430\\
                 & $\mu_c$ (fm$^{-1})$ &   0.70\\
                 & $\Delta$ (MeV)      & 181.10 \\[2ex]
                 & $\alpha_0$              & 2.118 \\
                 & $\Lambda_0~$(fm$^{-1}$) & 0.113 \\
OGE              & $\mu_0~$(MeV)        & 36.976\\
                 & $\hat{r}_0~$(MeV~fm) & 28.170\\
\end{tabular}
\end{ruledtabular}
\end{table}

\begin{figure}[ht]
\epsfxsize=1.4in \epsfbox{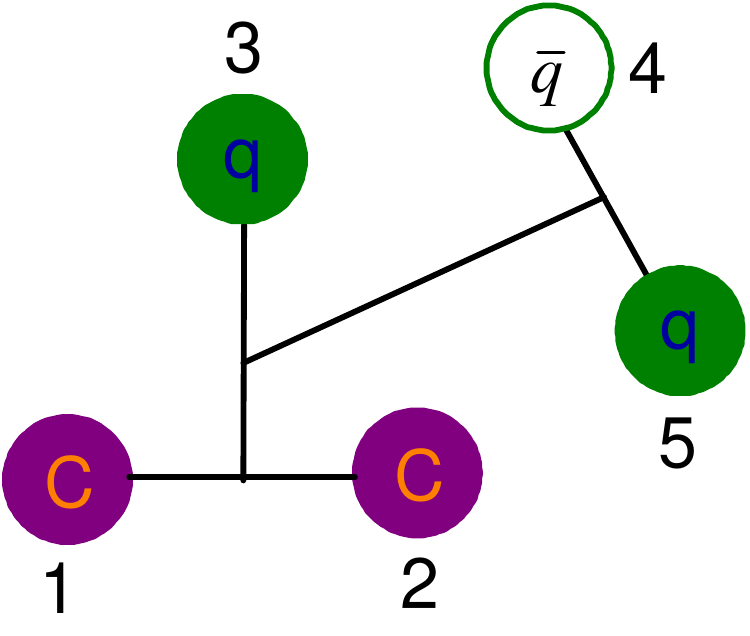}
\caption{The configuration of $ccqq\bar{q}$ $(q=u~or~d)$ pentaquarks. Two charmed quarks are in one cluster and coupled firstly.} \label{PS1}
\end{figure}

\begin{figure}[ht]
\epsfxsize=1.4in \epsfbox{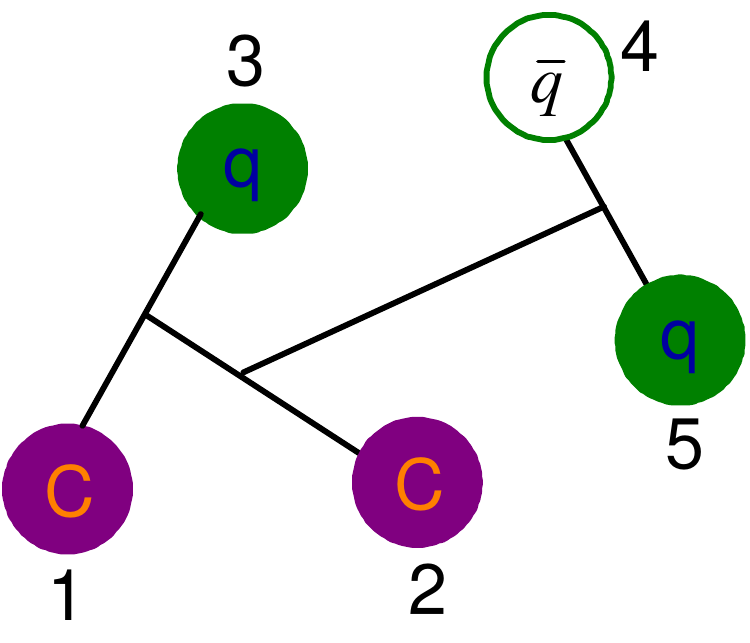}
\caption{The configuration of $ccqq\bar{q}$ $(q=u~or~d)$ pentaquarks. Two charmed quarks are in one cluster with the light- and heavy-quark coupled firstly.} \label{PS2}
\end{figure}

\begin{figure}[ht]
\epsfxsize=1.4in \epsfbox{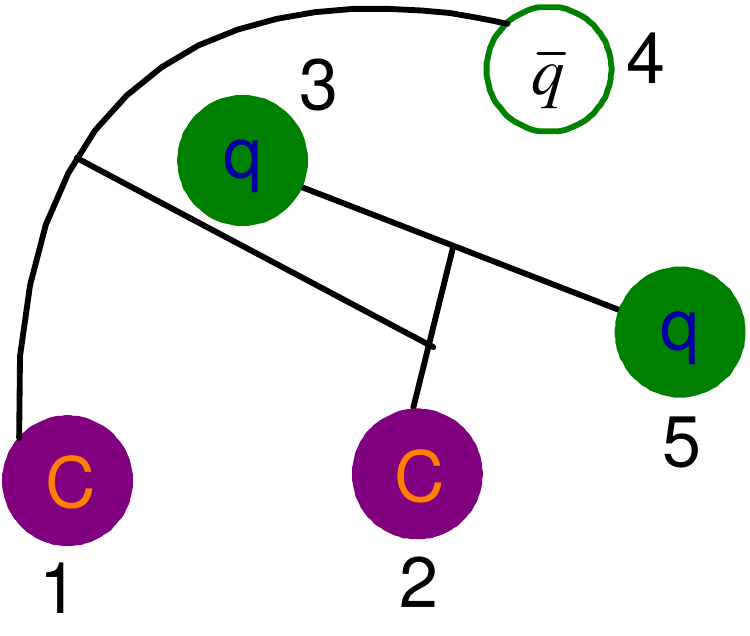}
\caption{The configuration of $ccqq\bar{q}$ $(q=u~or~d)$ pentaquarks. The two heavy quarks are divided into two clusters with light-quarks coupled firstly.} \label{PS3}
\end{figure}

\begin{figure}[ht]
\epsfxsize=1.4in \epsfbox{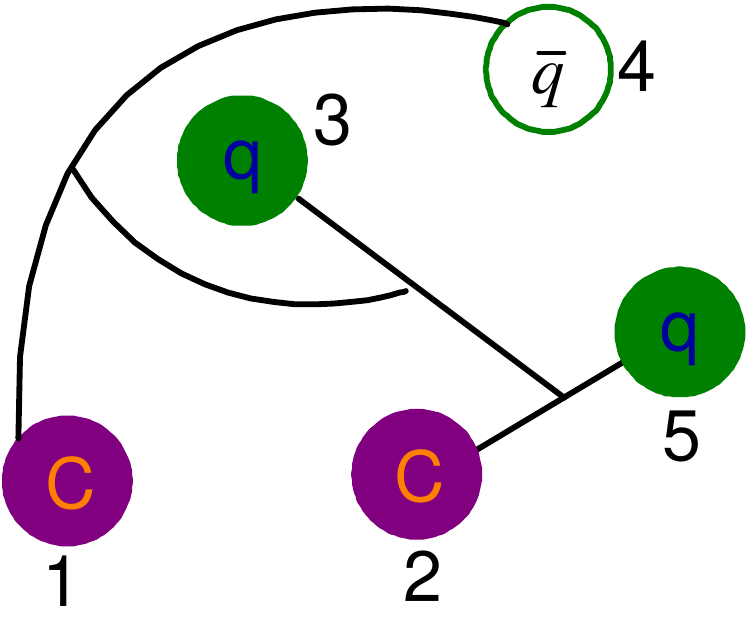}
\caption{The configuration of $ccqq\bar{q}$ $(q=u~or~d)$ pentaquarks. The two heavy quarks are divided into two clusters with the light- and heavy-quark coupled firstly.} \label{PS4}
\end{figure}

There are four sets of baryon-meson configurations for $ccqq\bar{q}~(q=u~or~d)$ systems\footnote{Note that the diquark-diquark-antiquark configuration is not considered herein because it goes beyond the scope of this work.}, and they are shown in Figs.~\ref{PS1} to~\ref{PS4}. Moreover, the anti-symmetry property in these identical fermion systems is necessary; however, due to the asymmetry between light and heavy quarks, the two charmed quarks can be coupled first within a 3-quark cluster as shown in Fig.~\ref{PS1}. Therefore, the antisymmetry operator for the $ccq\bar{q}q$ pentaquark system is 
\begin{equation}
{\cal{A}}_1 = 1-(35) \,. \label{EE1}
\end{equation}
Figure~\ref{PS2} shows a different arrangement in the 3-quark cluster with two heavy quarks included. In this case, the antisymmetry operator is given by
\begin{equation}
{\cal{A}}_2 = 1-(12)-(35)+(12)(35) \,. \label{EE2}
\end{equation}

The cases in which the two charm quarks are separated in different clusters are also considered and shown in Figs.~\ref{PS3} and~\ref{PS4}. When the two light quarks are coupled as in Fig.~\ref{PS3}, the antisymmetry operator is
\begin{equation}
{\cal{A}}_3 = 1-(12) \,, \label{EE3}
\end{equation}
whereas the last configuration, shown in Fig.~\ref{PS4}, has the same antisymmetry operator of Eq.~(\ref{EE2}); this is to say
\begin{equation}
{\cal{A}}_4 = {\cal{A}}_2 \,. \label{EE4}
\end{equation}

The pentaquark wave function is a product of four terms: color, flavor, spin and space wave functions. Concerning the color degrees-of-freedom, multiquark systems have richer structure than the conventional mesons and baryons. For instance, the $5$-quark wave function must be colorless but the way of reaching this condition can be done through either a color-singlet or a hidden-color channel, or both. The authors of Refs.~\cite{Harvey:1980rva, Vijande:2009kj} assert that it is enough to consider the color singlet channel when all possible excited states of a system are included. However, a more economical way of computing is considering both; the color singlet wave function:
\begin{align}
\label{Color1}
\chi^{n c}_1 &= \frac{1}{\sqrt{18}}(rgb-rbg+gbr-grb+brg-bgr) \times \nonumber \\
&
\times (\bar r r+\bar gg+\bar bb) \,,
\end{align}
where n=1-4 is a label for each quark configuration shown in Figs.~\ref{PS1} to~\ref{PS4}, respectively (it is of the same meaning for spin, flavor and space wave functions). In other words, they have a common form but with different quark sequence: 123;45, 132;45, 352;41 and 253;41. When computing matrix elements, one should switch the last three cases into the first one. The hidden-color channel is given by:
\begin{align}
\label{Color2}
\chi^{n c}_k &= \frac{1}{\sqrt{8}}(\chi^{n k}_{3,1}\chi_{2,8}-\chi^{n k}_{3,2}\chi_{2,7}-\chi^{n k}_{3,3}\chi_{2,6}+\chi^{n k}_{3,4}\chi_{2,5} \nonumber \\
& +\chi^{n k}_{3,5}\chi_{2,4}-\chi^{n k}_{3,6}\chi_{2,3}-\chi^{n k}_{3,7}\chi_{2,2}+\chi^{n k}_{3,8}\chi_{2,1}) \,,
\end{align}
where $k=2\,(3)$ is an index which stands for the symmetric (anti-symmetric) configuration of two quarks in the $3$-quark cluster. All color configurations have been used herein, as in the case of the $P_c^+$ $(P_b^+)$ hidden-charm (-bottom) pentaquarks studied in Refs.~\cite{Yang:2015bmv, Yang:2018oqd}.

According to the SU(2) symmetry in isospin space, the flavor wave functions for the clusters mentioned above are given by:
\begin{align}
B^3_{11}  &= uuc \,,  \,\,\,\,\, B^3_{1-1} = ddc \,, \\ 
B^4_{11}  &= ucu \,,  \,\,\,\,\, B^4_{1-1} = dcd \,, \\ 
B^3_{10}  &= \frac{1}{\sqrt{2}}(ud+du)c \,, \\ 
B^4_{10}  &= \frac{1}{\sqrt{2}}(ucd+dcu) \,, \\ 
B^3_{00} &= \frac{1}{\sqrt{2}}(ud-du)c \,, \\
B^4_{00} &= \frac{1}{\sqrt{2}}(ucd-dcu) \,, \\ 
B^1_{\frac12,\frac12} &= ccu \,,   \,\,\,\,\, B^1_{\frac12,-\frac12} = ccd \,, \\
B^2_{\frac12,\frac12} &= cuc \,,   \,\,\,\,\, B^2_{\frac12,-\frac12} = cdc \,, \\
M_{\frac12,\frac12} &= \bar{d}c \,,  \, M_{\frac12,-\frac12} = -\bar{u}c \,, \\ 
M_{11} &= \bar{d}u \,,   \,\,\,\,\, M_{1-1} = -\bar{u}d \,, \\ 
M_{10} &= -\frac{1}{\sqrt{2}}(\bar{u}u-\bar{d}d) \,, \\ 
M_{00} &= -\frac{1}{\sqrt{2}}(\bar{u}u+\bar{d}d) \,,
\end{align}
where the superscript of the flavor wave functions of 3-quark clusters stand for the number of each pentaquark configuration. Consequently, the flavor wave-functions for the 5-quark system with isospin $I=1/2$ or $3/2$ are
\begin{align}
&
\chi_{\frac12,\frac12}^{n f1}(5) = \sqrt{\frac{2}{3}} B^n_{11} M_{\frac12,-\frac12} -\sqrt{\frac{1}{3}} B^n_{10} M_{\frac12,\frac12} \,, \\
&
\chi_{\frac12,\frac12}^{n f2}(5) = B^n_{00} M_{\frac12,\frac12} \,,\\
&
\chi_{\frac12,\frac12}^{n f3}(5) = B^n_{\frac12,\frac12} M_{00} \,, \\
&
\chi_{\frac12,\frac12}^{n f4}(5) = -\sqrt{\frac{2}{3}} B^n_{\frac12,-\frac12} M_{11} +\sqrt{\frac{1}{3}} B^n_{\frac12,\frac12} M_{10}  \,, \\
& 
\chi_{\frac32,\frac32}^{n f1}(5) = B^n_{\frac12,\frac12} M_ {1,1}\,, \\
& 
\chi_{\frac32,\frac32}^{n f2}(5) = B^n_{1,1} M_{\frac12,\frac12} \,, 
%
%\chi^f_1  & = & \sqrt{\frac{2}{3}} B_{11} M_{\frac12,-\frac12}
 %-\sqrt{\frac{1}{3}} B_{10} M_{\frac12,\frac12}, \nonumber \\
%\chi^f_2  & = & B_{00} M_{\frac12,\frac12}, \\
%\chi^f_3  & = & B_{\frac12,\frac12}^1 M_{00}, \nonumber \\
%\chi^f_4  & = & B_{\frac12,\frac12}^2 M_{00}, \nonumber
\end{align}
where the third component of the isospin is set to be equal to the total one without loss of generality, because there is no interaction term in the Hamiltonian that can distinguish such component.

We consider herein 5-quark systems with total spin ranging from $1/2$ to $5/2$. Our Hamiltonian does not have any spin-orbital coupling dependent potential, and thus we can assume that the spin wave function has its third component equal to the total one, without loss of generality:
\begin{align}
\label{Spin}
\chi_{\frac12,\frac12}^{n \sigma 1}(5) &= \sqrt{\frac{1}{6}} \chi_{\frac32,-\frac12}^{n \sigma}(3) \chi_{11}^{\sigma}
-\sqrt{\frac{1}{3}} \chi_{\frac32,\frac12}^{n \sigma}(3) \chi_{10}^{\sigma} \nonumber \\
&
+\sqrt{\frac{1}{2}} \chi_{\frac32,\frac32}^{n \sigma}(3) \chi_{1-1}^{\sigma} \,, \\
\chi_{\frac12,\frac12}^{n \sigma 2}(5) &= \sqrt{\frac{1}{3}} \chi_{\frac12,\frac12}^{n \sigma 1}(3) \chi_{10}^{\sigma} -\sqrt{\frac{2}{3}} \chi_{\frac12,-\frac12}^{n \sigma 1}(3) \chi_{11}^{\sigma} \,, \\
\chi_{\frac12,\frac12}^{n \sigma 3}(5) &= \sqrt{\frac{1}{3}} \chi_{\frac12,\frac12}^{n \sigma 2}(3) \chi_{10}^{\sigma} - \sqrt{\frac{2}{3}} \chi_{\frac12,-\frac12}^{n \sigma 2}(3) \chi_{11}^{\sigma} \,, \\
\chi_{\frac12,\frac12}^{n \sigma 4}(5) &= \chi_{\frac12,\frac12}^{n \sigma 1}(3) \chi_{00}^{\sigma} \,, \\
\chi_{\frac12,\frac12}^{n \sigma 5}(5) &= \chi_{\frac12,\frac12}^{n \sigma 2}(3) \chi_{00}^{\sigma} \,,
\end{align}
for $S=1/2$, and
\begin{align}
\chi_{\frac32,\frac32}^{n \sigma 1}(5) &= \sqrt{\frac{3}{5}}
\chi_{\frac32,\frac32}^{n \sigma}(3) \chi_{10}^{\sigma} -\sqrt{\frac{2}{5}} \chi_{\frac32,\frac12}^{n \sigma}(3) \chi_{11}^{\sigma} \,, \\
\chi_{\frac32,\frac32}^{n \sigma 2}(5) &= \chi_{\frac32,\frac32}^{n \sigma}(3) \chi_{00}^{\sigma} \,, \\
\chi_{\frac32,\frac32}^{n \sigma 3}(5) &= \chi_{\frac12,\frac12}^{n \sigma 1}(3) \chi_{11}^{\sigma} \,, \\
\chi_{\frac32,\frac32}^{n \sigma 4}(5) &= \chi_{\frac12,\frac12}^{n \sigma 2}(3) \chi_{11}^{\sigma} \,,
\end{align}
for $S=3/2$, and
\begin{align}
\chi_{\frac52,\frac52}^{n \sigma 1}(5) &= \chi_{\frac32,\frac32}^{n \sigma}(3) \chi_{11}^{\sigma} \,,
\end{align}
for $S=5/2$. These expressions can be obtained easily using SU(2) algebra and considering the 3-quark and quark-antiquark clusters separately. They were derived in Ref.~\cite{Yang:2015bmv} for the hidden-charm pentaquarks.

The complex Schr\"odinger-like 5-body bound state equation is solved using the Rayleigh-Ritz variational principle, which is one of the most extended tools to solve eigenvalue problems due to its simplicity and flexibility. However, it is of great importance how to choose the basis on which to expand the wave function. The spatial wave function of a $5$-quark system is written as follows:
\begin{align}
\label{eq:WFexp}
&
\psi_{LM_L}=[ [ [ \phi_{n_1l_1}(\mbox{\boldmath $\rho$} e^{i\theta})\phi_{n_2l_2}(\mbox{\boldmath $\lambda$} e^{i\theta})]_{l} \phi_{n_3l_3}(\mbox{\boldmath $r$} e^{i\theta}) ]_{l^{\prime}} \nonumber\\
&
\hspace*{1.60cm}  \phi_{n_4l_4}(\mbox{\boldmath $R$} e^{i\theta}) ]_{LM_L} \,.
\end{align}
Taking the first pentaquark configuration shown in Fig.~\ref{PS1} as an example\footnote{The other three configurations are similar and differ only in the arrangement of quark sequence.}, the internal Jacobi coordinates are defined as
\begin{align}
{\mbox{\boldmath $\rho$}} &= {\mbox{\boldmath $x$}}_1-{\mbox{\boldmath $x$}}_2 \,, \\
{\mbox{\boldmath $\lambda$}} &= {\mbox{\boldmath $x$}}_3 - (\frac{{m_1\mbox{\boldmath $x$}}_1+{m_2\mbox{\boldmath $x$}}_2}{m_1+m_2}) \,,  \\
{\mbox{\boldmath $r$}} &= {\mbox{\boldmath $x$}}_4-{\mbox{\boldmath $x$}}_5 \,, \\
{\mbox{\boldmath $R$}} &= \left(\frac{{m_1\mbox{\boldmath $x$}}_1+{m_2\mbox{\boldmath $x$}}_2 + {m_3\mbox{\boldmath $x$}}_3}{m_1+m_2+m_3}\right) \nonumber \\
&
- \left(\frac{{m_4\mbox{\boldmath $x$}}_4+{m_5\mbox{\boldmath $x$}}_5}{m_4+m_5}\right) \,.
\end{align}
This choice is convenient because, on one hand, the center-of-mass kinetic term $T_{CM}$ can be completely eliminated for a nonrelativistic system and, in the other hand, the spatial wave functions related with the relative motions between quarks can be also extended to the complex scaling.

In order to make the calculation tractable, even for complicated interactions, we replace the orbital wave functions, $\phi$'s in Eq.~\eqref{eq:WFexp}, by a superposition of infinitesimally-shifted Gaussians (ISG)~\cite{Hiyama:2003cu}:
\begin{align}
&
\phi_{nlm}(\vec{r} e^{i\theta}\,) = N_{nl} (r e^{i\theta})^{l} e^{-\nu_{n} (r e^{i\theta})^2} Y_{lm}(\hat{r}) \nonumber \\
&
= N_{nl} \lim_{\varepsilon\to 0} \frac{1}{(\nu_{n}\varepsilon)^l} \sum_{k=1}^{k_{\rm
max}} C_{lm,k} e^{-\nu_{n}(\vec{r} e^{i\theta}-\varepsilon \vec{D}_{lm,k})^{2}} \,.
\end{align}
where the limit $\varepsilon\to 0$ must be carried out after the matrix elements have been calculated analytically. This new set of basis functions makes the calculation of 5-body matrix elements easier without the laborious Racah algebra. Moreover, all the advantages of using Gau\ss ians remain with the new basis functions.

Finally, in order to fulfill the Pauli principle, the complete antisymmetry complex wave-function is written as
\begin{equation}
\label{TPs}
\Psi_{JM,i,j,k} (\theta)=\sum_{n=1}^4 {\cal A}_n \left[ \left[ \psi^n_{L} (\theta) \chi^{n \sigma_i}_{S}(5) \right]_{JM} \chi^{n f_j}_I \chi^{n c}_k \right] \,,
\end{equation}
where ${\cal A}_n$ is the antisymmetry operator of the 5-quark system and their expressions are shown in Eqs.~(\ref{EE1}) to~(\ref{EE4}). This is needed because we have constructed an antisymmetric wave function for only two quarks of the 3-quark cluster, the remaining (anti-)quarks of the system have been added to the wave function by simply considering the appropriate Clebsch-Gordan coefficients.

%%%%%%%%%%%%%%%%%%%%%%%%%%%%%%%%%%%%%%%%%%%%%%%%%%%%%%%%%%%%%%%%%%%%%%%%%%%%%%%%

\section{Results}
\label{sec:results}

\begin{table*}[!t]
\caption{\label{GCC1} All possible channels for open-charm pentaquark systems with $J^P=1/2^-$.}
\begin{ruledtabular}
\begin{tabular}{cccccc}
& & \multicolumn{2}{c}{$I=\frac{1}{2}$} & \multicolumn{2}{c}{$I=\frac{3}{2}$} \\[2ex]
$J^P$~~&~Index~ & $\chi_J^{n \sigma_i}$;~$\chi_I^{n f_j}$;~$\chi_k^{n c}$; & Channel~~ & $\chi_J^{n \sigma_i}$;~$\chi_I^{n f_j}$;~$\chi_k^{n c}$; & Channel~~ \\
&&$[i; ~j; ~k;~n]$& &$[i; ~j; ~k;~n]$&  \\[2ex]
$\frac{1}{2}^-$ & 1  & $[4; ~3; ~1; ~1,2]$   & $(\Xi_{cc} \eta)^1$ & $[4; ~2; ~1; ~1,2]$   & $(\Xi_{cc} \pi)^1$ \\
&  2 & $[4,5; ~3; ~2,3; ~1,2]$ & $(\Xi_{cc} \eta)^8$ & $[4,5; ~2; ~2,3; ~1,2]$  & $(\Xi_{cc} \pi)^8$ \\
&  3 & $[2; ~3; ~1; ~1,2]$   & $(\Xi_{cc} \omega)^1$  & $[2; ~2; ~1; ~1,2]$   & $(\Xi_{cc} \rho)^1$ \\
&  4 & $[2,3; ~3; ~2,3; ~1,2]$ & $(\Xi_{cc} \omega)^8$ & $[2,3; ~2; ~2,3; ~1,2]$   & $(\Xi_{cc} \rho)^8$   \\
&  5 & $[4; ~4; ~1; ~1,2]$     & $(\Xi_{cc} \pi)^1$  & $[1; ~2; ~1; ~1,2]$   & $(\Xi^*_{cc} \rho)^1$ \\
&  6 & $[4,5; ~4; ~2,3; ~1,2]$ & $(\Xi_{cc} \pi)^8$  & $[1; ~2; ~3; ~1,2]$   & $(\Xi^*_{cc} \rho)^8$ \\
&  7 & $[2; ~4; ~1; ~1,2]$     & $(\Xi_{cc} \rho)^1$  & $[4; ~3; ~1; ~3,4]$   & $(\Sigma_c D)^1$ \\
&  8 & $[2,3; ~4; ~2,3; ~1,2]$ & $(\Xi_{cc} \rho)^8$  & $[4,5; ~3; ~2,3; ~3,4]$   & $(\Sigma_c D)^8$ \\
&  9 & $[1; ~3; ~1, ~1,2]$     & $(\Xi^*_{cc} \omega)^1$  & $[2; ~3; ~1; ~3,4]$   & $(\Sigma_c D^*)^1$ \\
& 10 & $[1; ~3; ~3, ~1,2]$ & $(\Xi^*_{cc} \omega)^8$  & $[2,3; ~3; ~2,3; ~3,4]$   & $(\Sigma_c D^*)^8$ \\
& 11 & $[1; ~4; ~1; ~1,2]$     & $(\Xi^*_{cc} \rho)^1$  & $[1; ~3; ~1; ~3,4]$   & $(\Sigma^*_c D^*)^1$ \\
& 12 & $[1; ~4; ~3; ~1,2]$ & $(\Xi^*_{cc} \rho)^8$  & $[1; ~3; ~3; ~3,4]$   & $(\Sigma^*_c D^*)^8$ \\
& 13 & $[5; ~2; ~1; ~3,4]$     & $(\Lambda_c D)^1$ & & \\
& 14 & $[4,5; ~2; ~2,3; ~3,4]$     & $(\Lambda_c D)^8$ & & \\
& 15 & $[3; ~2; ~1; ~3,4]$     & $(\Lambda_c D^*)^1$ & & \\
& 16 & $[2,3; ~2; ~2,3; ~3,4]$     & $(\Lambda_c D^*)^8$ & & \\
& 17 & $[4; ~1; ~1; ~3,4]$     & $(\Sigma_c D)^1$ & & \\
& 18 & $[4,5; ~1; ~2,3; ~3,4]$     & $(\Sigma_c D)^8$ & & \\
& 19 & $[2; ~1; ~1; ~3,4]$     & $(\Sigma_c D^*)^1$ & & \\
& 20 & $[2,3; ~1; ~2,3; ~3,4]$     & $(\Sigma_c D^*)^8$ & & \\
& 21 & $[1; ~1; ~1; ~3,4]$     & $(\Sigma^*_c D^*)^1$ & & \\
& 22 & $[1; ~1; ~3; ~3,4]$     & $(\Sigma^*_c D^*)^8$ & & \\[2ex]
\end{tabular}
\end{ruledtabular}
\end{table*}

\begin{table*}[!t]
\caption{\label{GCC2} All possible channels for open-charm pentaquark systems with $J^P=3/2^-$ and $5/2^-$.}
\begin{ruledtabular}
\begin{tabular}{cccccc}
& & \multicolumn{2}{c}{$I=\frac{1}{2}$} & \multicolumn{2}{c}{$I=\frac{3}{2}$} \\[2ex]
$J^P$~~&~Index~ & $\chi_J^{n \sigma_i}$;~$\chi_I^{n f_j}$;~$\chi_k^{n c}$; & Channel~~ & $\chi_J^{n \sigma_i}$;~$\chi_I^{n f_j}$;~$\chi_k^{n c}$; & Channel~~ \\
&&$[i; ~j; ~k;~n]$& &$[i; ~j; ~k;~n]$&  \\[2ex]
$\frac{3}{2}^-$ & 1  & $[3; ~3; ~1; ~1,2]$   & $(\Xi_{cc} \omega)^1$ & $[3; ~2; ~1; ~1,2]$ & $(\Xi_{cc} \rho)^1$\\
& 2  & $[3,4; ~3; ~2,3; ~1,2]$  & $(\Xi_{cc} \omega)^8$ & $[3,4; ~2; ~2,3; ~1,2]$  & $(\Xi_{cc} \rho)^8$ \\
& 3  & $[3; ~4; ~1; ~1,2]$     & $(\Xi_{cc} \rho)^1$  & $[2; ~2; ~1; ~1,2]$   & $(\Xi^*_{cc} \pi)^1$ \\
& 4  & $[3,4; ~4; ~2,3; ~1,2]$  & $(\Xi_{cc} \rho)^8$ & $[2; ~2; ~3; ~1,2]$ & $(\Xi^*_{cc} \pi)^8$ \\
& 5  & $[2; ~4; ~1; ~1,2]$     & $(\Xi^*_{cc} \pi)^1$  & $[1; ~2; ~1; ~1,2]$   & $(\Xi^*_{cc} \rho)^1$ \\
& 6  & $[2; ~4; ~3; ~1,2]$  & $(\Xi^*_{cc} \pi)^8$  & $[1; ~2; ~3; ~1,2]$   & $(\Xi^*_{cc} \rho)^8$  \\
& 7  & $[1; ~3; ~1; ~1,2]$     & $(\Xi^*_{cc} \omega)^1$  & $[3; ~3; ~1; ~3,4]$   & $(\Sigma_c D^*)^1$ \\
& 8  & $[1; ~3; ~3; ~1,2]$  & $(\Xi^*_{cc} \omega)^8$  & $[3,4; ~3; ~2,3; ~3,4]$   & $(\Sigma_c D^*)^8$  \\
& 9  & $[1; ~4; ~1; ~1,2]$   & $(\Xi^*_{cc} \rho)^1$ & $[2; ~3; ~1; ~3,4]$  & $(\Sigma^*_c D)^1$ \\
& 10 & $[1; ~4; ~3; ~1,2]$  & $(\Xi^*_{cc} \rho)^8$  & $[3; ~3; ~3; ~3,4]$   &  $(\Sigma^*_c D)^8$  \\
& 11  & $[4; ~2; ~1; ~3,4]$   & $(\Lambda_c D^*)^1$ & $[1; ~3; ~1; ~3,4]$  & $(\Sigma^*_c D^*)^1$ \\
& 12 & $[3,4; ~2; ~2,3; ~3,4]$  & $(\Lambda_c D^*)^8$  & $[1; ~3; ~3; ~3,4]$   &  $(\Sigma^*_c D^*)^8$  \\
& 13  & $[3; ~1; ~1; ~3,4]$   & $(\Sigma_c D^*)^1$ &   &  \\
& 14 & $[3,4; ~1; ~1; ~3,4]$  & $(\Sigma_c D^*)^8$  &   &  \\
& 15  & $[2; ~1; ~1; ~3,4]$   & $(\Sigma^*_c D)^1$ &   &  \\
& 16 & $[2; ~1; ~3; ~3,4]$  & $(\Sigma^*_c D)^8$  &   &  \\
& 17  & $[1; ~1; ~1; ~3,4]$   & $(\Sigma^*_c D^*)^1$ &   &  \\
& 18 & $[1; ~1; ~3; ~3,4]$  & $(\Sigma^*_c D^*)^8$  &   &  \\ [2ex]
$\frac{5}{2}^-$ & 1 & $[1; ~3; ~1; ~1,2]$  & $(\Xi^*_{cc} \omega)^1$ & $[1; ~2; ~1; ~1,2]$ & $(\Xi^*_{cc} \rho)^1$ \\
& 2  & $[1; ~3; ~3; ~1,2]$  & $(\Xi^*_{cc} \omega)^8$ & $[1; ~2; ~3; ~1,2]$  & $(\Xi^*_{cc} \rho)^8$ \\
& 3  & $[1; ~4; ~1; ~1,2]$   & $(\Xi^*_{cc} \rho)^1$   & $[1; ~3; ~1; ~3,4]$   & $(\Sigma^*_c D^*)^1$ \\
& 4  & $[1; ~4; ~3; ~1,2]$   & $(\Xi^*_{cc} \rho)^8$  & $[1; ~3; ~3; ~3,4]$ & $(\Sigma^*_c D^*)^8$ \\
& 5  & $[1; ~1; ~1; ~3,4]$   & $(\Sigma^*_c D^*)^1$  &  &  \\
& 6  & $[1; ~1; ~3; ~3,4]$   & $(\Sigma^*_c D^*)^8$  &  &  \\
\end{tabular}
\end{ruledtabular}
\end{table*}

In the present calculation, we investigate the possible lowest-lying and resonance states of the $ccqq\bar q$ $(q=u~or~d)$ pentaquark systems by taking into account the $(ccq)(\bar{q}q)$, $(cqc)(\bar{q}q)$, $(qqc)(\bar{q}c)$ and $(cqq)(\bar{q}c)$ configurations in which the considered baryons have always positive parity and mesons are either pseudoscalars $(J^P=0^-)$ or vectors $(1^-)$. This means that, in our approach, a pentaquark state with negative parity has $L=0$. In this case, we assume that the angular momenta $l_1$, $l_2$, $l_3$ and $l_4$, appearing in Eq.~\eqref{eq:WFexp}, are all equal to zero. Accordingly, the total angular momentum, $J$, coincides with the total spin, $S$, and can take values $1/2$, $3/2$ and $5/2$. The possible baryon-meson channels which are under consideration in the computation are listed in Tables~\ref{GCC1} and~\ref{GCC2}, they have been grouped according to total spin and parity $J^P$, and isospin $I$. The third and fifth columns of such Tables show the necessary basis combination in spin $(\chi^{n \sigma_i}_J)$, flavor $(\chi^{n f_j}_I)$, and color $(\chi^{n c}_k)$ degrees-of-freedom, along with the possible configurations $(n=1,\ldots,4)$ shown in Figs.~\ref{PS1},~\ref{PS2},~\ref{PS3} and~\ref{PS4}. The physical channels with color-singlet (labeled with the superindex $1$) and hidden-color (labeled with the superindex $8$) configurations are listed in the fourth and sixth columns of the same Tables.

Firstly, we perform a calculation of the lowest-lying doubly-charm pentaquarks with a rotated angle $\theta=0^o$. Tables~\ref{Gresult1},~\ref{Gresult3},~\ref{Gresult5},~\ref{Gresult6},~\ref{Gresult8} and~\ref{Gresult10} summarize our masses of the $ccqq\bar{q}$ systems with spin-parity $J^P=\frac{1}{2}^-$, $\frac{3}{2}^-$ and $\frac{5}{2}^-$, isospin $I=\frac{1}{2}$ and $\frac{3}{2}$, respectively. In each Table, the first and fourth column show the baryon-meson channel and also, in parenthesis, the experimental value of the noninteracting baryon-meson threshold; the second column refers to color-singlet (S), hidden-color (H) and coupled-channels (S+H) calculations; the third and fifth columns show the theoretical mass of the pentaquark state. All of these states are scattering ones and thus the corresponding binding energies are bigger than zero. However, also in real-range calculation $(\theta=0^\circ)$, Tables~\ref{Gresult2},~\ref{Gresult4},~\ref{Gresult7} and~\ref{Gresult9} show our findings about the possible existence of lowest-lying doubly-charm pentaquarks with quantum numbers $I(J^P)=\frac{1}{2}(\frac{1}{2}^-)$, $\frac{1}{2}(\frac{3}{2}^-)$, $\frac{3}{2}(\frac{1}{2}^-)$ and $\frac{3}{2}(\frac{3}{2}^-)$, respectively. In these tables, the first column shows the baryon-meson channel in which a bound state appears, it also indicates in parenthesis the experimental value of the noninteracting baryon-meson threshold; the second column refers to color-singlet (S), hidden-color (H) and coupled-channels (S+H) calculations; the third and fourth columns show the theoretical mass and binding energy of the pentaquark bound-state; and the fifth column presents the theoretical mass of the pentaquark state but re-scaled attending to the experimental baryon-meson threshold, this is in order to avoid theoretical uncertainties coming from the quark model prediction of the baryon and meson spectra.

\begin{figure}[!t]
\includegraphics[clip, trim={3.0cm 2.0cm 3.0cm 1.0cm}, width=0.45\textwidth]{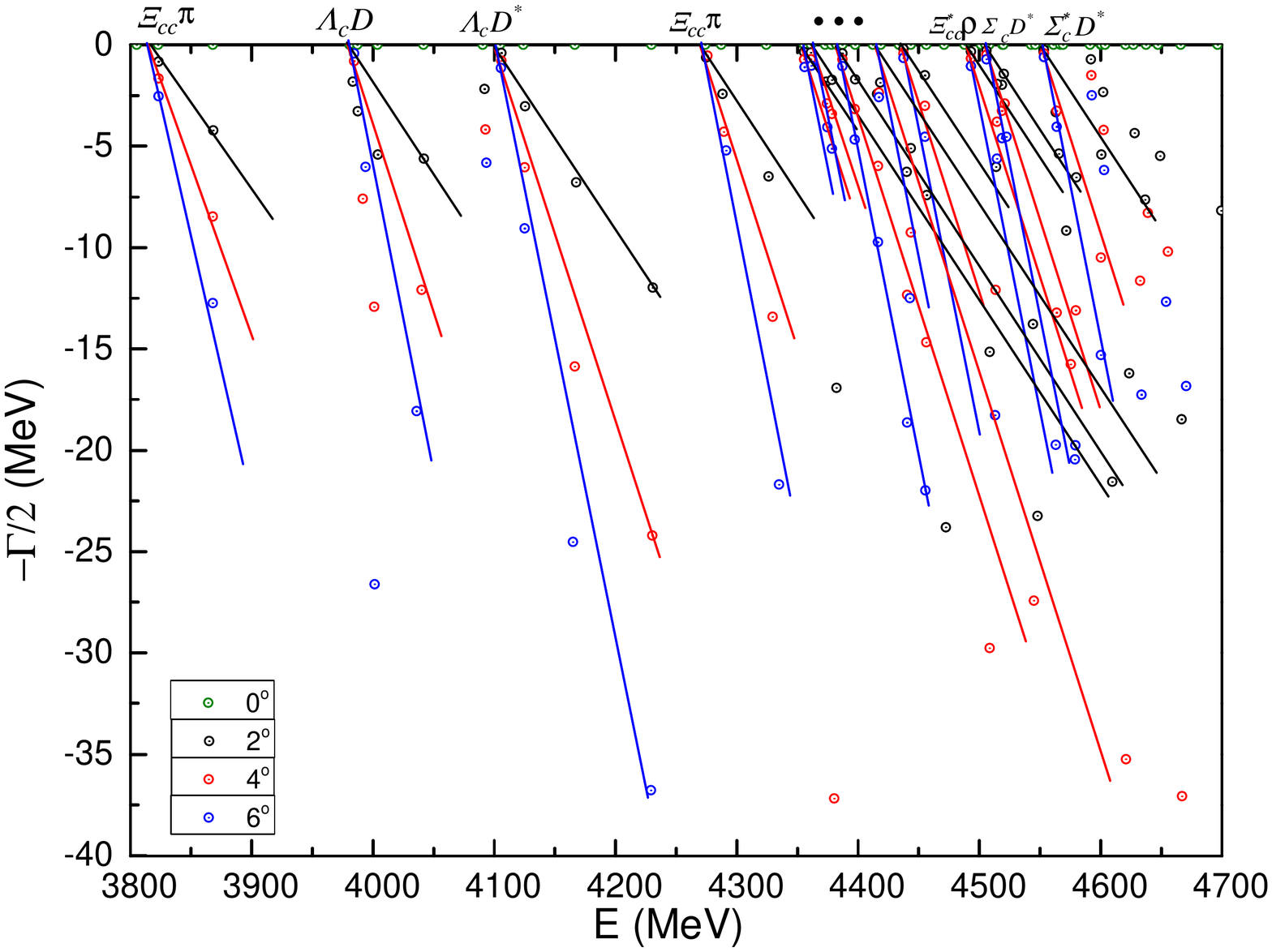} \\
\includegraphics[clip, trim={3.0cm 2.0cm 3.0cm 1.0cm}, width=0.45\textwidth]{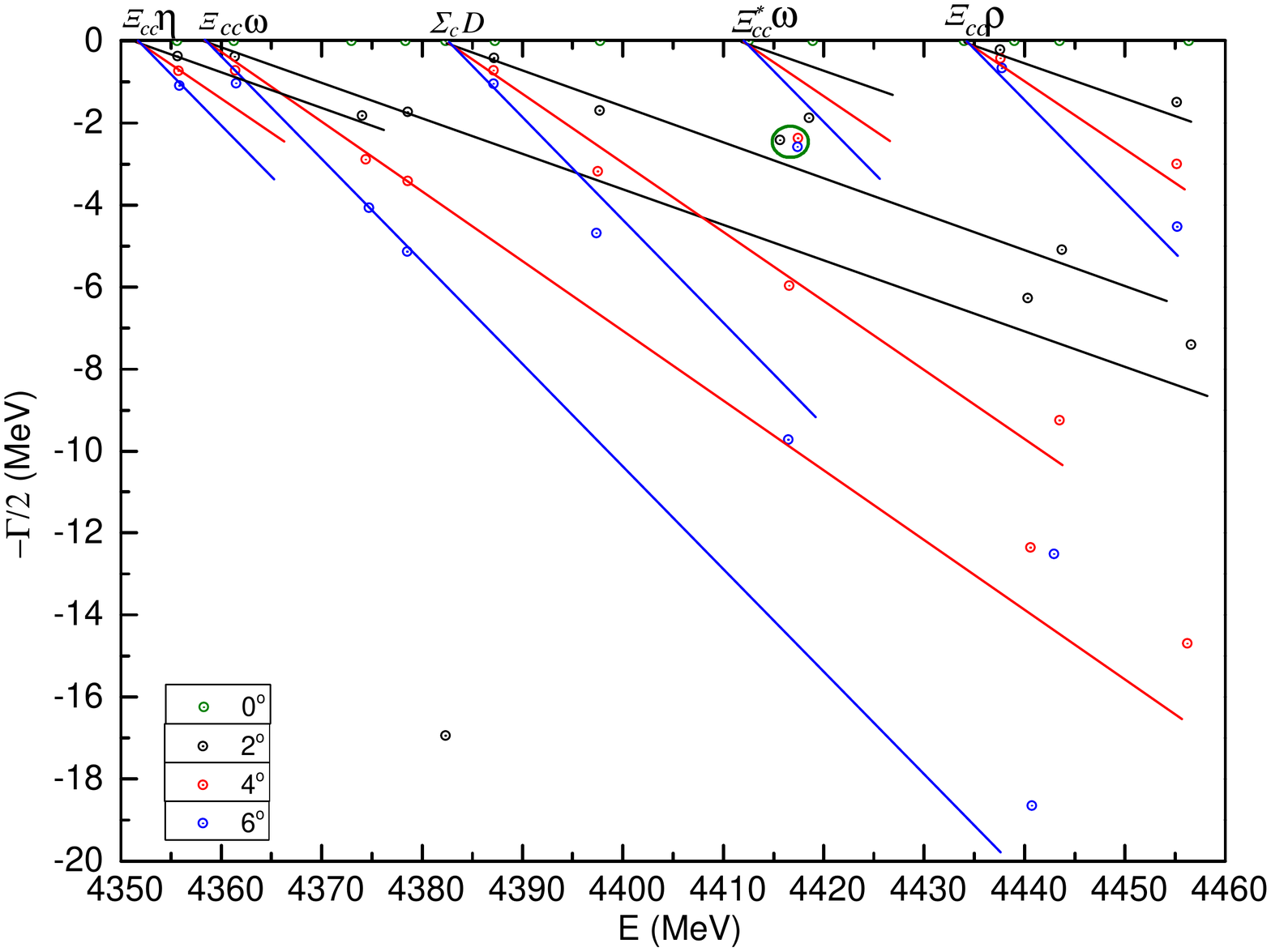}
\caption{\label{PP1} {\it Top panel:} Pentaquark's complex energies of coupled-channels calculation with quantum numbers $IJ^P=\frac12\frac12^-$ and for $\theta(^\circ)=0$ (green), $2$ (black), $4$ (red) and $6$ (blue). {\it Bottom panel:} Enlarged top panel, with real values of energy ranging from $4.35\,\text{GeV}$ to $4.46\,\text{GeV}$.} 
\end{figure}

\begin{figure}[!t]
\includegraphics[clip, trim={3.0cm 2.0cm 3.0cm 1.0cm}, width=0.45\textwidth]{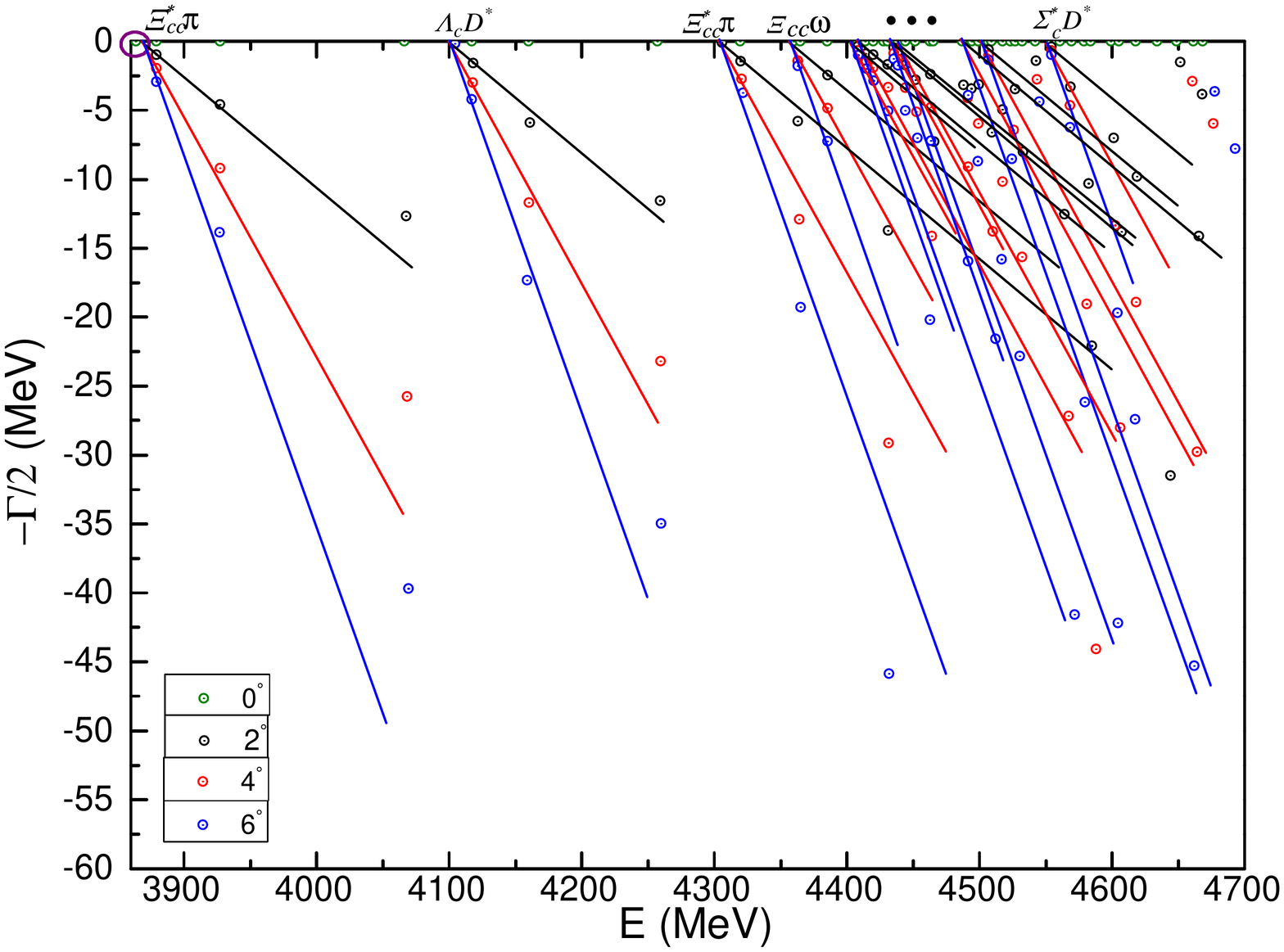} \\
\includegraphics[clip, trim={3.0cm 2.0cm 3.0cm 1.0cm}, width=0.45\textwidth]{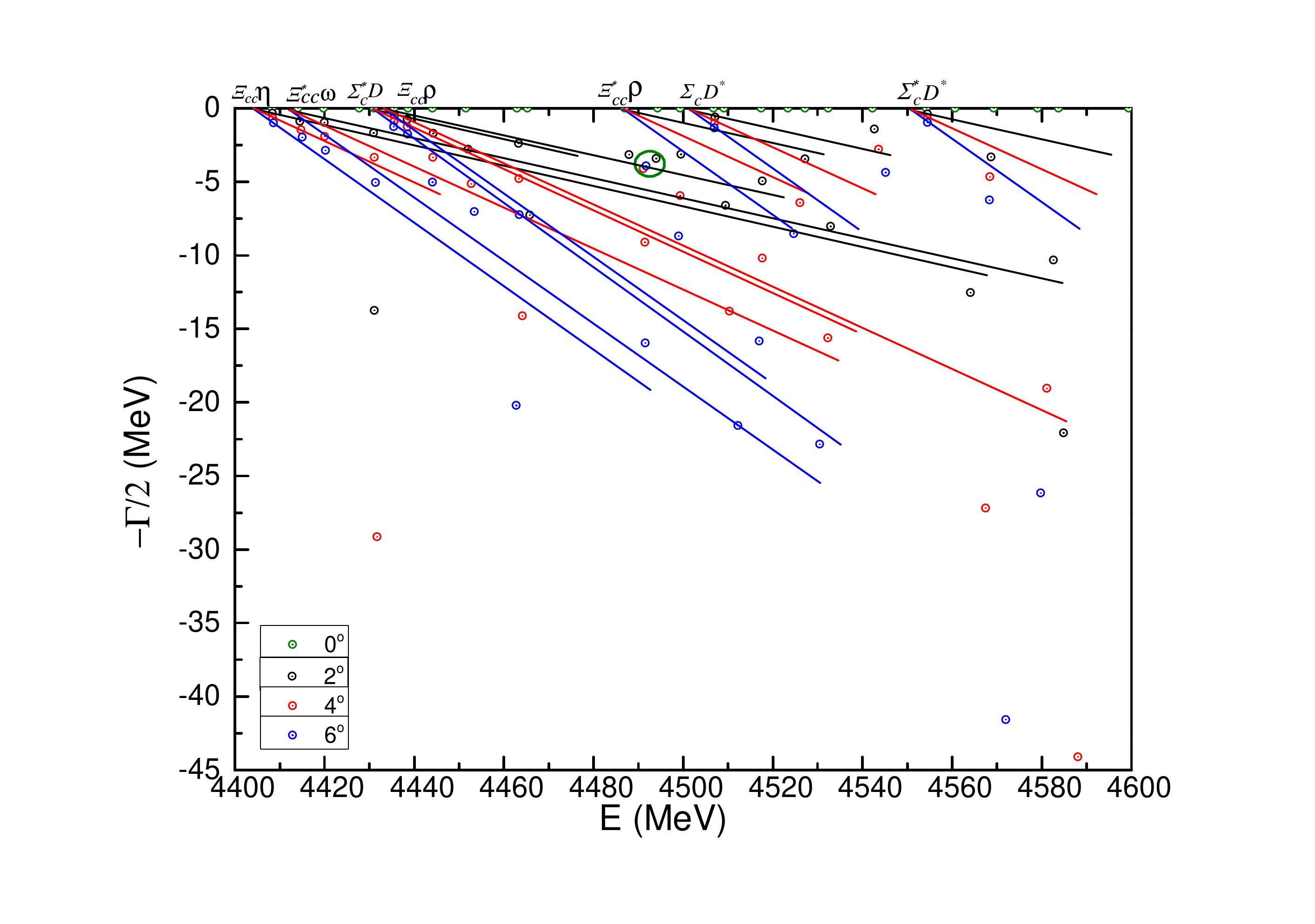}
\caption{\label{PP2} {\it Top panel:} Pentaquark's complex energies of coupled-channels calculation with quantum numbers $IJ^P=\frac12\frac32^-$ and for $\theta(^\circ)=0$ (green), $2$ (black), $4$ (red) and $6$ (blue). {\it Bottom panel:} Enlarged top panel, with real values of energy ranging from $4.40\,\text{GeV}$ to $4.60\,\text{GeV}$.} 
\end{figure}

\begin{figure}[!t]
\includegraphics[clip, trim={3.0cm 2.0cm 3.0cm 1.0cm}, width=0.45\textwidth]{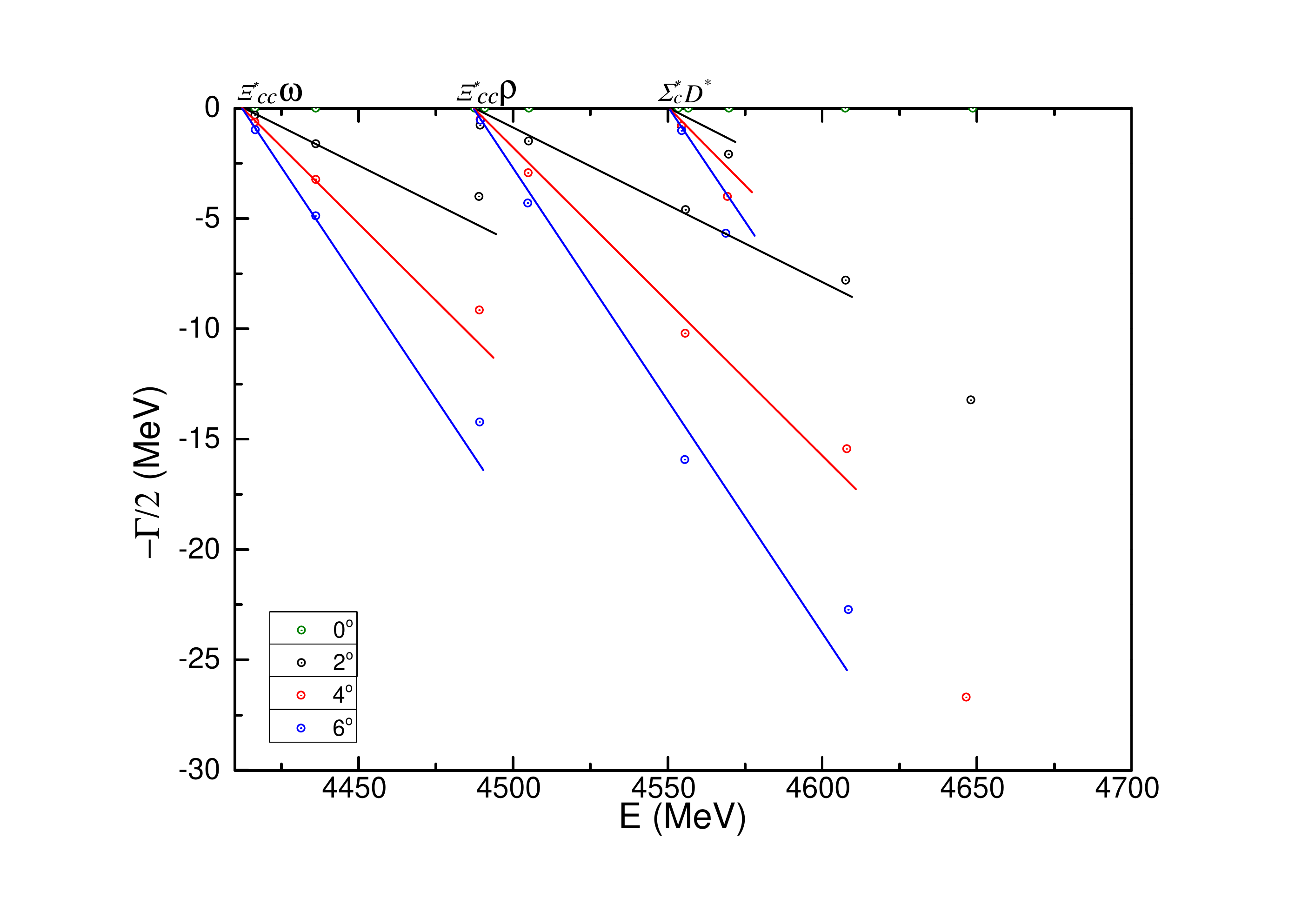}
\caption{\label{PP5} Pentaquark's complex energies of coupled-channels calculation with quantum numbers $IJ^P=\frac12\frac52^-$ and for $\theta(^\circ)=0$ (green), $2$ (black), $4$ (red) and $6$ (blue).} 
\end{figure}

\begin{figure}[!t]
\includegraphics[clip, trim={3.0cm 2.0cm 3.0cm 1.0cm}, width=0.45\textwidth]{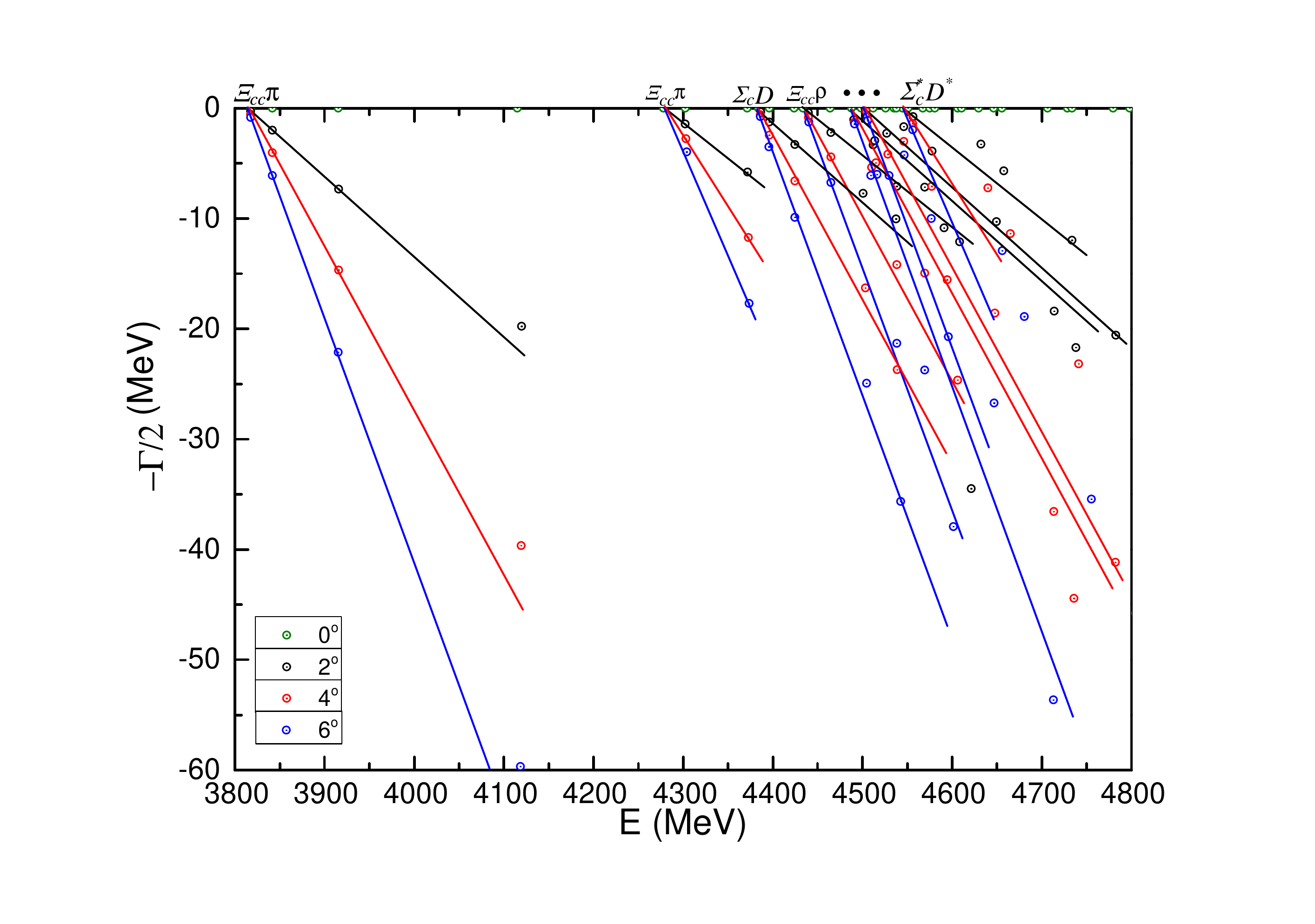} \\
\includegraphics[clip, trim={3.0cm 2.0cm 3.0cm 1.0cm}, width=0.45\textwidth]{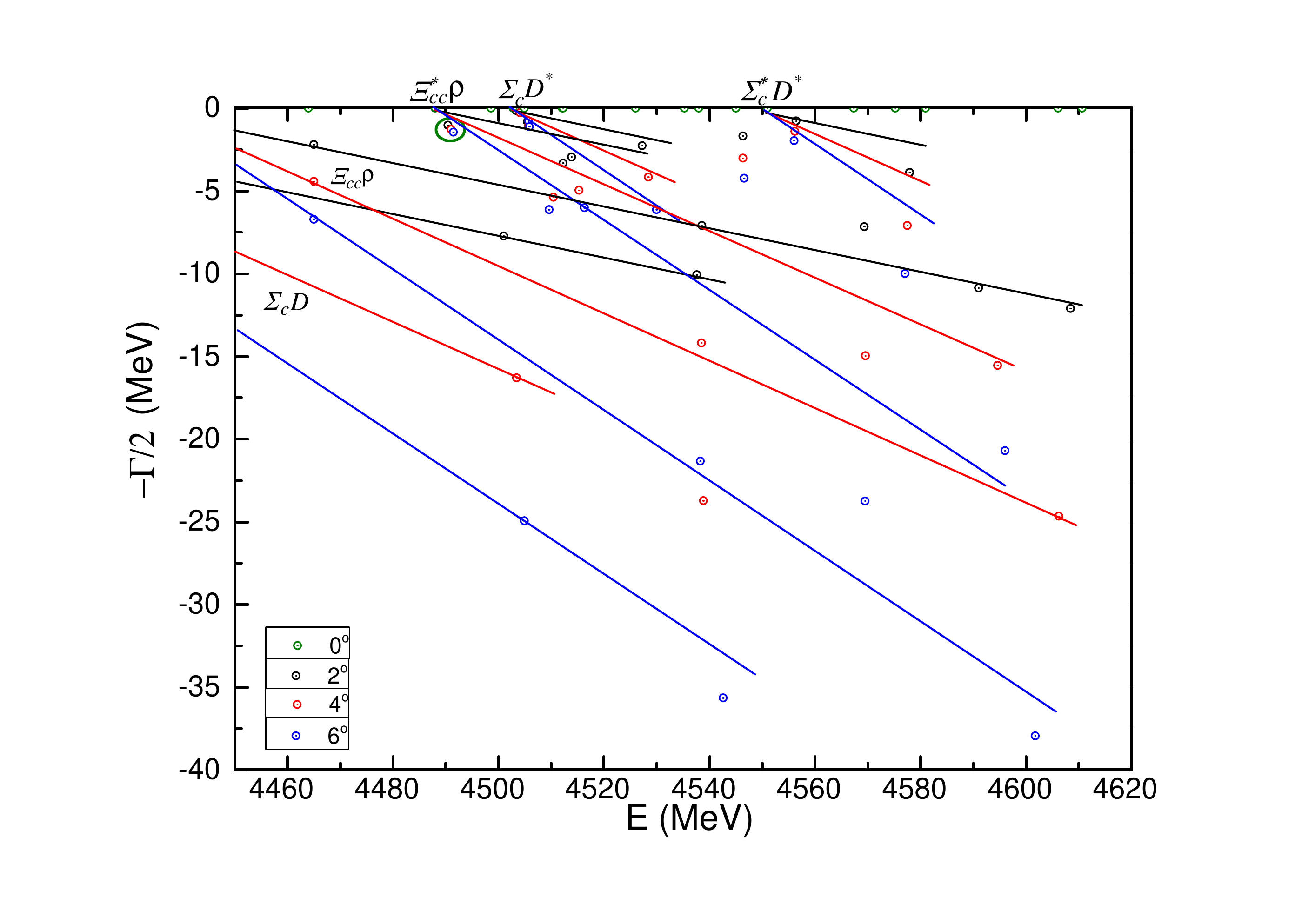}\\
\includegraphics[clip, trim={3.0cm 2.0cm 3.0cm 1.0cm}, width=0.45\textwidth]{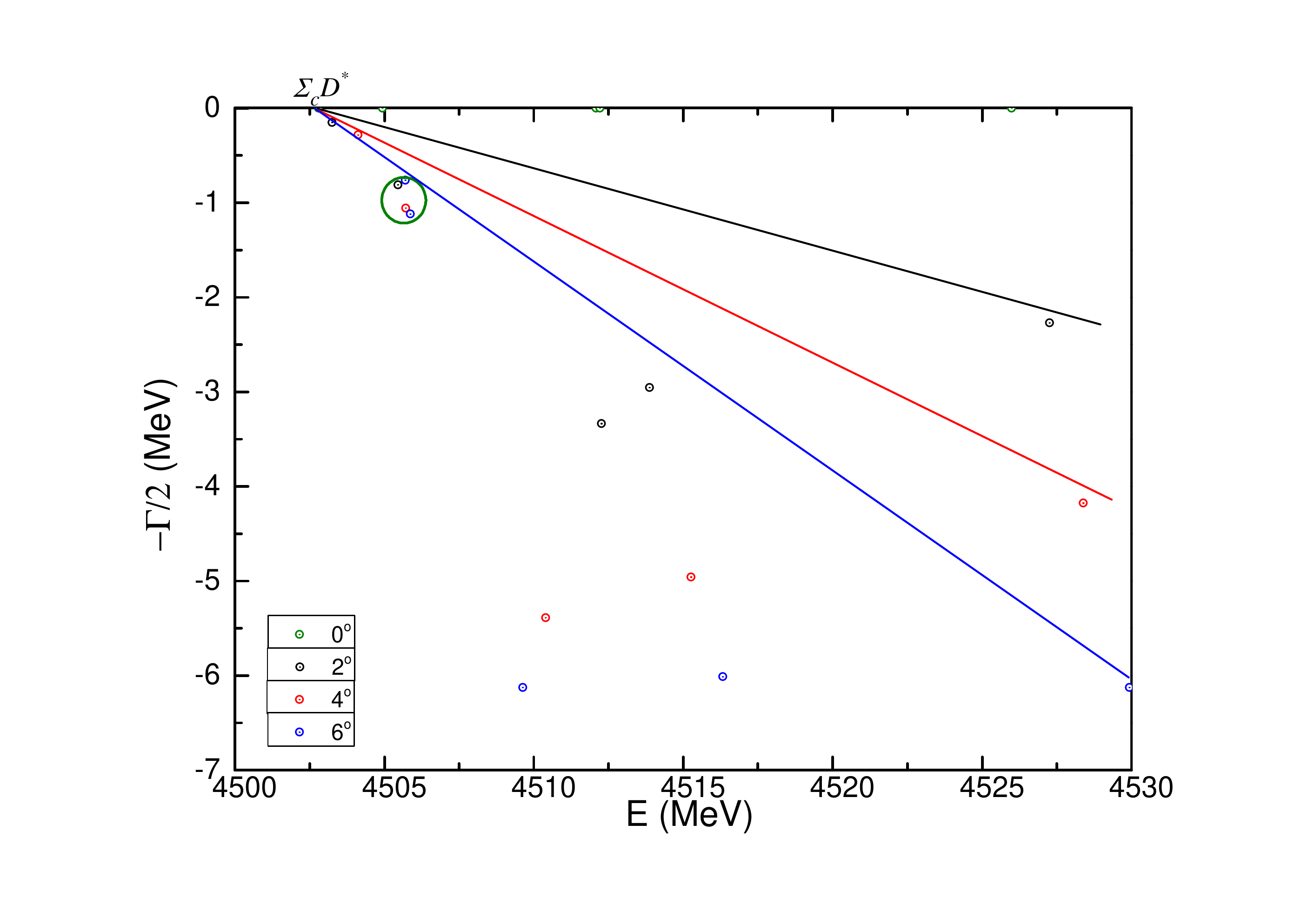}
\caption{\label{PP3} {\it Top panel:} Pentaquark's complex energies of coupled-channels calculation with quantum numbers $IJ^P=\frac32\frac12^-$ and for $\theta(^\circ)=0$ (green), $2$ (black), $4$ (red) and $6$ (blue). {\it Middle panel:} Enlarged top panel, with real values of energy ranging from $4.45\,\text{GeV}$ to $4.62\,\text{GeV}$. {\it Bottom panel:} Enlarged top panel, with real values of energy ranging from $4.50\,\text{GeV}$ to $4.53\,\text{GeV}$.} 
\end{figure}

\begin{figure}[!t]
\includegraphics[clip, trim={3.0cm 2.0cm 3.0cm 1.0cm}, width=0.45\textwidth]{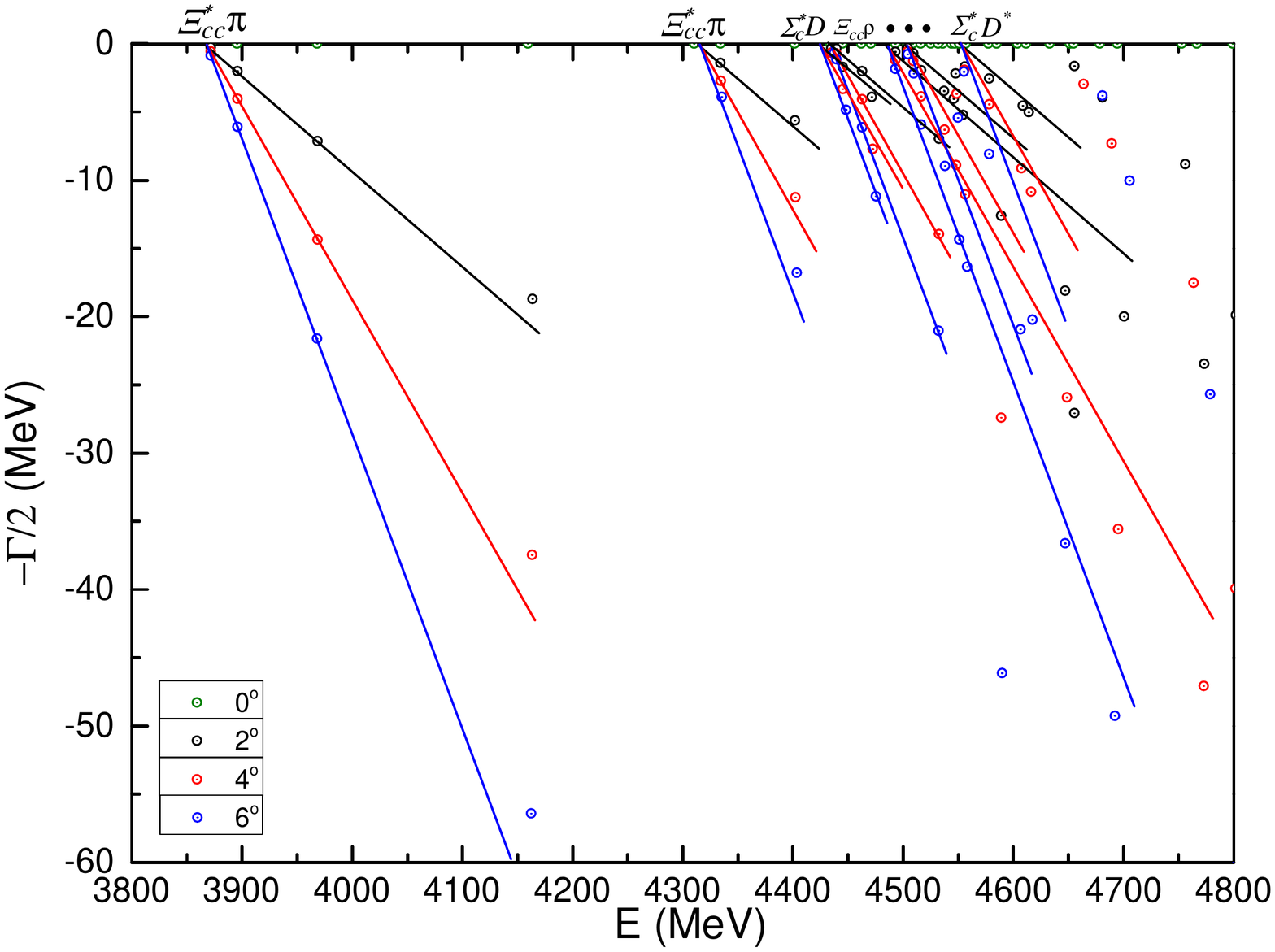} \\
\includegraphics[clip, trim={3.0cm 2.0cm 3.0cm 1.0cm}, width=0.45\textwidth]{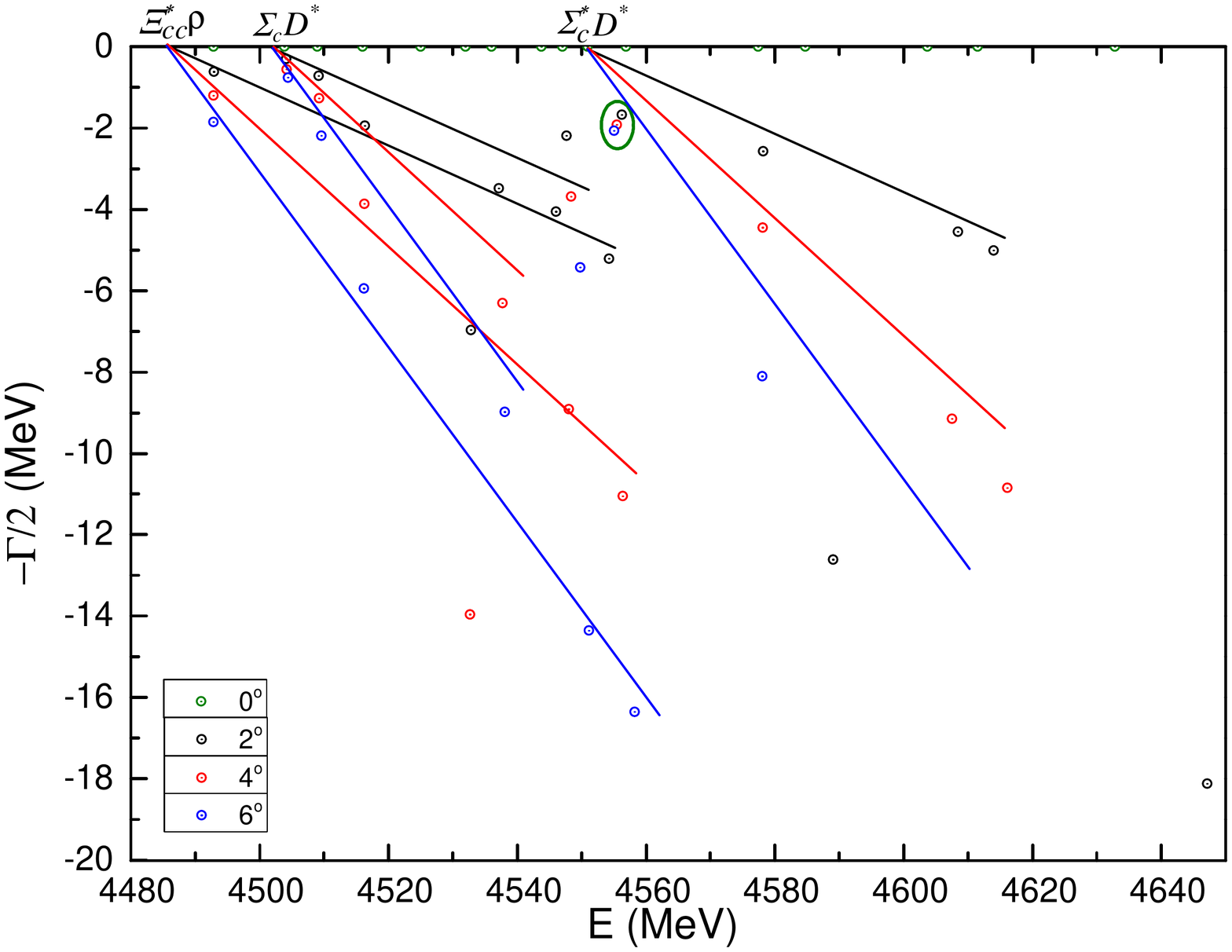}
\caption{\label{PP10} {\it Top panel:} Pentaquark's complex energies of coupled-channels calculation with quantum numbers $IJ^P=\frac32\frac32^-$ and for $\theta(^\circ)=0$ (green), $2$ (black), $4$ (red) and $6$ (blue). {\it Bottom panel:} Enlarged top panel, with real values of energy ranging from $4.48\,\text{GeV}$ to $4.65\,\text{GeV}$.} 
\end{figure}

\begin{figure}[!t]
\includegraphics[clip, trim={3.0cm 2.0cm 3.0cm 1.0cm}, width=0.45\textwidth]{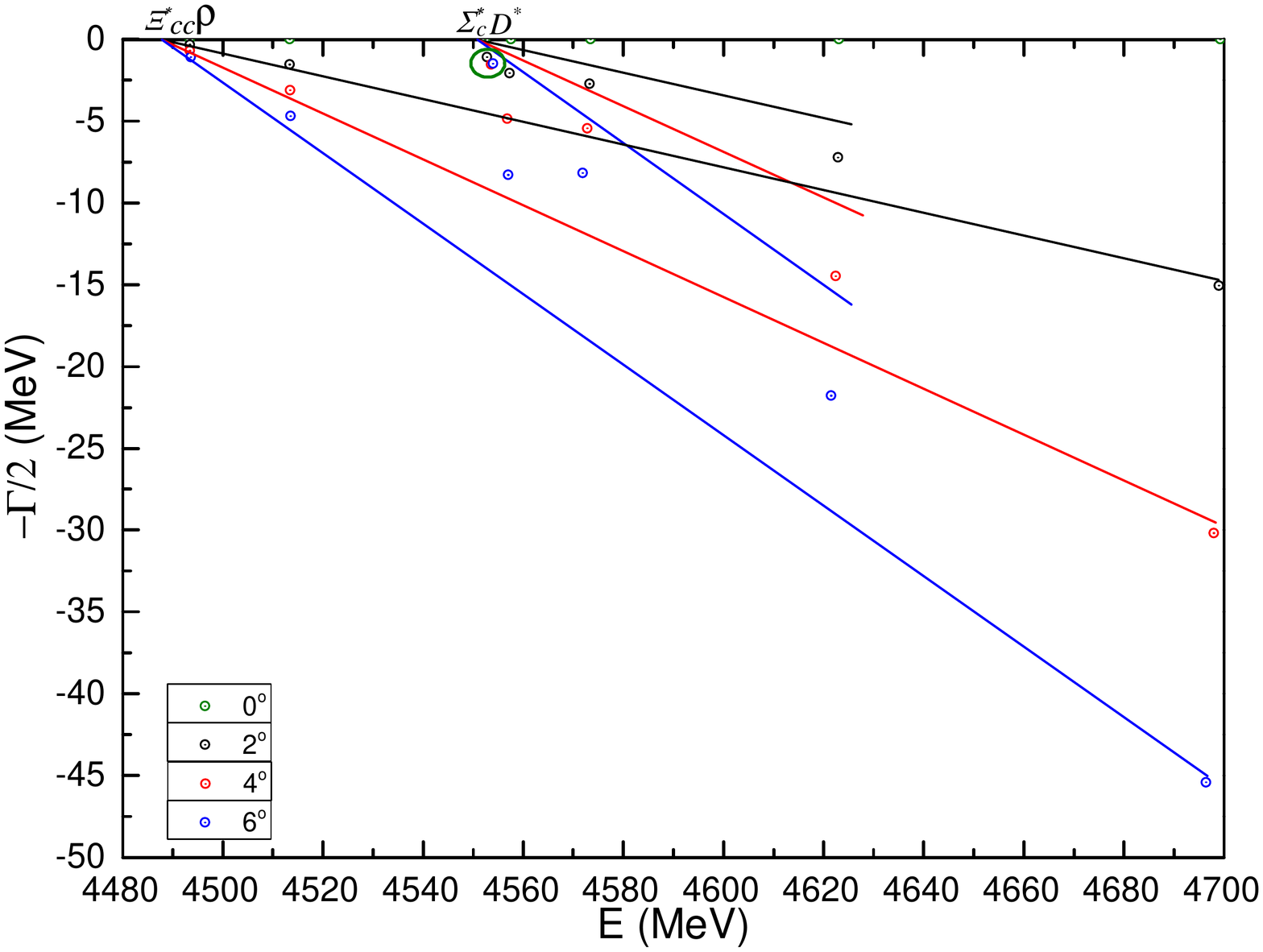}
\caption{\label{PP11} Pentaquark's complex energies of coupled-channels calculation with quantum numbers $IJ^P=\frac32\frac52^-$ and for $\theta(^\circ)=0$ (green), $2$ (black), $4$ (red) and $6$ (blue).} 
\end{figure}

In addition to the study sketched briefly in the last paragraph, we use the mentioned complex scaling method (CSM) to investigate the nature of a given pentaquark state in coupled-channels calculation. There exist (resonance) poles for pentaquark states with quantum numbers $I(J^P)=\frac{1}{2}(\frac{1}{2}^-)$, $\frac{1}{2}(\frac{3}{2}^-)$, $\frac{3}{2}(\frac{1}{2}^-)$, $\frac{3}{2}(\frac{3}{2}^-)$ and $\frac{3}{2}(\frac{5}{2}^-)$. No resonance state is found in the present work with total spin $J^P=\frac{5}{2}^-$ and isospin $I=\frac12$. As for those possible resonance states, their complex energies (masses and widths) are established in Figs.~\ref{PP1} to~\ref{PP11}. Moreover, Table~\ref{Rsum} summarized our theoretical findings of these possible bound and resonance states.

We proceed now to describe in detail our theoretical findings:

\begin{table}[!t]
\caption{\label{Gresult1} The lowest eigen-eneries of doubly-charm pentaquarks with $I(J^P)=\frac12(\frac12^-)$, and the rotated angle $\theta=0^\circ$. (unit: MeV) }
\begin{ruledtabular}
\begin{tabular}{lcccc}
Channel   & Color & $M$ & Channel & $M$ \\[2ex]
$\Xi_{cc}\eta$ & S   & $4351$ & $\Xi_{cc}\omega$  & $4358$ \\
$(4065)$         & H   & $4787$ & $(4300)$ & $4608$ \\
                  & S+H & $4351$ & & $4358$ \\[2ex]
$\Xi_{cc}\pi$ & S   & $3812$ & $\Xi_{cc}\rho$  & $4434$ \\
$(3657)$         & H   & $4620$ & $(4293)$ & $4613$ \\
                  & S+H & $3812$ & & $4434$ \\[2ex]
$\Xi^*_{cc}\omega$ & S   & $4412$ & $\Xi^*_{cc}\rho$  & $4488$ \\
$(4403)$         & H   & $4568$ & $(4396)$ & $4576$ \\
                  & S+H & $4412$ & & $4488$ \\[2ex]
$\Lambda_c D$ & S   & $3981$ & $\Sigma^*_c D^*$  & $4551$ \\
$(4155)$         & H   & $4299$ & $(4527)$ & $4779$ \\
                  & S+H & $3981$ & & $4551$ \\[2ex]
$\Sigma_c D$ & S   & $4384$ & $\Sigma_c D^*$  & $4503$ \\
$(4324)$         & H   & $4701$ & $(4462)$ & $4691$ \\
                  & S+H & $4384$ & & $4503$ \\
%     ~~~~~Mixed (singlet)  & 10195 & &  \\
%     ~~~~~Mixed (full)  & 10195 & & \\  \hline\hline
\end{tabular}
\end{ruledtabular}
\end{table}

\begin{table}[!t]
\caption{\label{Gresult2} The lowest eigen-eneries of $\Lambda_c D^*$ with $I(J^P)=\frac12(\frac12^-)$, and the rotated angle $\theta=0^\circ$. (unit: MeV) }
\begin{ruledtabular}
\begin{tabular}{lcccc}
Channel   & Color & $M$ & $E_B$ & $M'$ \\[2ex]
$\Lambda_c D^*$ & S   & $4098$ & $-2$  & $4291$ \\
$(4293)$         & H   & $4312$ & $+212$ & $4505$ \\
                  & S+H & $4098$ & $-2$  & $4291$ \\
%     ~~~~~Mixed (singlet)  & 10195 & &  \\
%     ~~~~~Mixed (full)  & 10195 & & \\  \hline\hline
\end{tabular}
\end{ruledtabular}
\end{table}

{\bf The $\bm{I(J^P)=\frac12(\frac12^-)}$ channel:} Among all the possible baryon-meson channels: $\Xi_{cc} \eta$, $\Xi_{cc} \omega$, $\Xi_{cc} \pi$, $\Xi_{cc} \rho$, $\Xi^*_{cc} \omega$, $\Xi^*_{cc} \rho$, $\Lambda_c D$, $\Lambda_c D^*$, $\Sigma_c D$, $\Sigma_c D^*$ and $\Sigma^*_c D^*$, only $\Lambda_c D^*$ is possibly bound in real-range calculation with a binding energy $E_B=-2\,\text{MeV}$ and its modified mass is $4291\,\text{MeV}$. One can clearly see in Table~\ref{Gresult2} that the coupling between color-singlet and hidden-color channels is quite weak. However, after a coupled-channels calculation for all of these possible channels in complex-range with a rotated angle $\theta$ varied from $0^\circ$ to $6^\circ$, one possible $\Sigma_c D$ resonance state is obtained. 

The distribution of complex energies with quantum numbers $I(J^P)=\frac12(\frac12^-)$ are shown in Fig.~\ref{PP1}. The green dots on the positive real-axis are the masses of coupled-channels calculation with $\theta=0^\circ$. Meanwhile, black, red and blue dots are for those with $\theta=2^\circ$, $4^\circ$ and $6^\circ$, respectively. Generally, they are aligned along the threshold lines with the same color. If we focus on, {\it e.g.}, $\Xi_{cc} \pi$ channel, whose lowest theoretical mass is $3812\,\text{MeV}$, the nature of scattering state is clearly identified because the obtained poles always move along the cut lines when the scaling angle $\theta$ changes. This feature is also observed for the other channels: $\Lambda_c D$, $\Lambda_c D^*$, $\Xi^*_{cc} \rho$, $\Sigma_c D^*$ and $\Sigma^*_c D^*$. Note, too, that the radial excited state of $\Xi_{cc} \pi$ is also obtained, as shown in Fig.~\ref{PP1}.

An important feature to highlight here is the following. The bound state of $\Lambda_c D^*$, with a mass of $4291\,\text{MeV}$, is pushed above its threshold within the coupled-channels calculation. In Fig.~\ref{PP1}, one can see that the pole of $\Lambda_c D^*$ is always going down with larger values of $\theta$. Since we are working with a finite Fock-space, some numerical noise is found in the high energy region, from $4.6\,\text{GeV}$. This issue can be settled with a large number Gaussian basis; however, such higher energies are not interesting for the scope of this work.

The top panel of Fig.~\ref{PP1} also shows a dense distribution of $\Xi_{cc} \eta$, $\Xi_{cc} \omega$, $\Sigma_c D$, $\Xi^*_{cc} \omega$ and $\Xi_{cc} \rho$ states in the energy region $4.35-4.46\,\text{GeV}$; for this reason, the bottom panel shows an enlarged version of it which concentrates on $[4.35-4.46]\,\text{GeV}$. One can see that the calculated complex energies fall mostly into the kind of continuum states, except a possible resonance pole whose mass and width are $\sim 4416\,\text{MeV}$ and $\sim 4.8\,\text{MeV}$, respectively. In the same figure, there are three almost overlaping points, circled in green, which correspond to the CSM calculation with $\theta=2^\circ$, $4^\circ$ and $6^\circ$. These points correspond to a resonance state which is above the threshold of $\Sigma_c D$. After a mass shift according to this channel, the re-scaled mass for the $\Sigma_c D$ resonance is $4356\,\text{MeV}$, with a width of $4.8\,\text{MeV}$. The nature of this resonance state, $\Sigma_c D(4356)$, is very similar to the $P^+_c(4312)$ hidden-charm pentaquark observed by the LHCb Collaboration, {\it i.e.} its quantum numbers $IJ^P$ are $\frac12 \frac12^-$ which are the ones preferred for the $P^+_c(4312)$~\cite{Yang:2015bmv, MZL190311560, JH190311872, CWX190401296, CJX190400872}; moreover, its 5-quark configuration is identified with a molecular state of $\Sigma_c \bar{D}$ with mass and width $4311.9\pm 0.7^{+6.8}_{-0.6}\,\text{MeV}$ and $9.8\pm 2.7^{+3.7}_{-4.5}\,\text{MeV}$, respectively. Hence, this new resonance state is expected to be identified in near future high-energy physics experiments.

%%%%%%%%%%

\begin{table}[!t]
\caption{\label{Gresult3} The lowest eigen-eneries of doubly-charm pentaquarks with $I(J^P)=\frac12(\frac32^-)$, and the rotated angle $\theta=0^\circ$. (unit: MeV)}
\begin{ruledtabular}
\begin{tabular}{lcccc}
Color   & Channel & $M$ & Channel & $M$ \\[2ex]
S   & $\Xi_{cc}\omega$ & $4358$ & $\Xi_{cc}\rho$  & $4434$ \\
H   & $(4300)$         & $4619$ & $(4293)$ & $4648$ \\
S+H &  & $4358$ & & $4434$ \\[2ex]
$\Xi^*_{cc}\pi$ & S   & $3866$ & $\Xi^*_{cc}\omega$ & $4412$ \\
$(3760)$        & H   & $4671$ & $(4403)$           & $4614$ \\
                & S+H & $3866$ &                    & $4412$ \\[2ex]
$\Xi^*_{cc}\rho$ & S   & $4488$ & $\Lambda_c D^*$ & $4100$ \\
$(4396)$         & H   & $4641$ & $(4293)$        & $4284$ \\
                 & S+H & $4488$ &                 & $4100$ \\[2ex]
$\Sigma_c D^*$    & S   & $4503$ & $\Sigma^*_c D$ & $4432$ \\
$(4462)$          & H   & $4689$ & $(4389)$       & $4702$ \\
                  & S+H & $4503$ &                & $4432$ \\[2ex]
$\Sigma^*_c D^*$  & S   & $4551$ & & \\
$(4527)$          & H   & $4729$ & & \\
                  & S+H & $4551$ & & \\
%     ~~~~~Mixed (singlet)  & 10195 & &  \\
%     ~~~~~Mixed (full)  & 10195 & & \\  \hline\hline
\end{tabular}
\end{ruledtabular}
\end{table}

\begin{table}[!t]
\caption{\label{Gresult4} The lowest eigen-eneries of $\Sigma_c D^*$ with $I(J^P)=\frac12(\frac32^-)$, and the rotated angle $\theta=0^\circ$. (unit: MeV) }
\begin{ruledtabular}
\begin{tabular}{lcccc}
Channel        & Color & $M$    & $E_B$  & $M'$ \\[2ex]
$\Sigma_c D^*$ & S     & $4503$ & $0$    & $4462$ \\
$(4462)$       & H     & $4689$ & $+186$ & $4648$ \\
               & S+H   & $4502$ & $-1$   & $4461$ \\
%     ~~~~~Mixed (singlet)  & 10195 & &  \\
%     ~~~~~Mixed (full)  & 10195 & & \\  \hline\hline
\end{tabular}
\end{ruledtabular}
\end{table}

{\bf The $\bm{I(J^P)=\frac12(\frac32^-)}$ channel:} The baryon-meson channels studied in this case are $\Xi^{(*)}_{cc} \omega$, $\Xi^{(*)}_{cc} \rho$, $\Xi^*_{cc} \pi$, $\Lambda_c D^*$ and $\Sigma^{(*)}_c D^{(*)}$, and Table~\ref{Gresult3} shows our findings with $\theta=0^\circ$. The definition of each column is the same as that in Table~\ref{Gresult1} of the $I(J^P)=\frac12 (\frac12^-)$ case. No bound state is found in these channels; however, a loosely bound one of $\Sigma_c D^*$ with a binding energy of $E=-1\,\text{MeV}$ could be obtained, as shown in Table~\ref{Gresult4}. For the possible bound state of $\Sigma_c D^*(4461)$, hidden-color channel helps a little in forming the baryon-meson molecular state. 

When the rotated angle $\theta$ is varied from $0^\circ$ to $6^\circ$ in coupled-channels calculation, several interesting results are observed. In Fig.~\ref{PP2}, the possible channels are mostly scattering states moving along their corresponding cut lines. Besides, there is a $\Xi^*_{cc} \pi(3757)$ bound state circled with purple in the real axis. Its binding energy is $E=-3\,\text{MeV}$ when compared with the threshold's theoretical value, $3866\,\text{MeV}$ in Table~\ref{Gresult3}. Therefore, after a mass shift with respect to the experimental value $3760\,\text{MeV}$, the modified mass is $3757\,\text{MeV}$. Consequently, the coupled-channels calculation results in a $\Xi^*_{cc} \pi(3757)$ bound state with $I(J^P)=\frac12 (\frac32^-)$, and  it is also expected to be observed in future experiment.

Similar to the case of $\Lambda_c D^*(4291)$ with $IJ^P=\frac12 \frac12^-$, the original bound state of $\Sigma_c D^*(4461)$ turns to be a scattering one due to interacting effects of lower channels $(\Xi^*_{cc} \pi,~\Lambda_c D^*,~\Xi^{(*)}_{cc} \omega,~\Xi_{cc} \eta,~\Sigma^*_c D~and~\Xi^{(*)}_{cc} \rho)$. The nature of $\Sigma_c D^*$ scattering state can be identified clearly in Fig.~\ref{PP2} where the corresponding calculated poles ($E\sim 4.5~GeV$ in real axis) go always down when increasing the rotated angle, $\theta$.

An enlarged figure for the energy region $4.4-4.6\,\text{GeV}$ is shown in the bottom panel of Fig.~\ref{PP2}. A resonance state is obtained and surrounded by a green circle (three calculated results of different $\theta$ are almost unchanged inside of it). The resonance's mass and width are about $4492\,\text{MeV}$ and $8.0\,\text{MeV}$, respectively. Due to this pole is above two almost degenerate scattering states of $\Sigma^*_c D$ and $\Xi_{cc} \rho$ whose theoretical thresholds are $4432\,\text{MeV}$ and $4434\,\text{MeV}$, in present work, the obtained resonance state is preferred to be identified as a molecular state of $\Sigma_c^* D$. Hence, after a mass shift according to $\Sigma_c^* D(4389)$ with $\Delta_{\text{threshold}}=43\,\text{MeV}$, the obtained resonance state has a mass of $E=4449\,\text{MeV}$ and a width of $\Gamma=8.0\,\text{MeV}$ respectively. Note again that there is also a significant similarity between $\Sigma^*_c D(4449)$ and the hidden-charm $P^+_c(4457)$ state. The later one is explained as $\Sigma_c \bar{D}^*$ molecular state with quantum numbers $IJ^P=\frac12 \frac32^-$~\cite{Yang:2015bmv, MZL190311560, JH190311872, CWX190401296, CJX190400872}, and its experimental mass and width is $4457.3\pm 0.6^{+4.1}_{-1.7}\,\text{MeV}$ and $6.4\pm 2.0^{+5.7}_{-1.9}\,\text{MeV}$, respectively. The nature of our candidate $\Sigma^*_c D(4449)$ molecular state is deserved to be investigated in future experimental facilities.

%%%%%%%%%%

\begin{table}[!t]
\caption{\label{Gresult5} The lowest eigen-eneries of doubly-charm pentaquarks with $I(J^P)=\frac12(\frac52^-)$, and the rotated angle $\theta=0^\circ$. (unit: MeV) }
\begin{ruledtabular}
\begin{tabular}{lcccc}
Channel   & Color & $M$ & Channel & $M$ \\[2ex]
$\Xi^*_{cc}\omega$ & S   & $4412$ & $\Xi^*_{cc}\rho$ & $4488$ \\
$(4403)$           & H   & $4683$ & $(4396)$         & $4741$ \\
                   & S+H & $4412$ &                  & $4488$ \\[2ex]
$\Sigma^*_c D^*$ & S   & $4551$ & & \\
$(4527)$         & H   & $4655$ & & \\
                 & S+H & $4551$ & & \\
%     ~~~~~Mixed (singlet)  & 10195 & &  \\
%     ~~~~~Mixed (full)  & 10195 & & \\  \hline\hline
\end{tabular}
\end{ruledtabular}
\end{table}

{\bf The $\bm{I(J^P)=\frac12(\frac52^-)}$ channel:} Table~\ref{Gresult5} lists the masses of possible states in the channels $\Xi^*_{cc} \omega$, $\Xi^*_{cc} \rho$ and $\Sigma^*_c D^*$, taking into account singlet-color, hidden-color and their coupling. The real-range calculation with rotated angle $\theta=0^\circ$ does not provide bound states. In a further complex-scaling study within coupled-channels calculation, neither bound nor resonance states are obtained. In Fig.~\ref{PP5}, the continuum states of $\Xi^*_{cc} \omega$, $\Xi^*_{cc} \rho$ and $\Sigma^*_c D^*$ are shown and they basically fall along the corresponding cut lines.

%%%%%%%%%%

\begin{table}[!t]
\caption{\label{Gresult6} The lowest eigen-eneries of doubly-charm pentaquarks with $I(J^P)=\frac32(\frac12^-)$, and the rotated angle $\theta=0^\circ$. (unit: MeV) }
\begin{ruledtabular}
\begin{tabular}{lcccc}
Channel   & Color & $M$ & Channel & $M$ \\[2ex]
$\Xi_{cc}\pi$ & S   & $3812$ & $\Xi_{cc}\rho$ & $4434$ \\
$(3657)$      & H   & $4682$ & $(4293)$       & $4685$ \\
              & S+H & $3812$ &                & $4434$ \\[2ex]
$\Xi^*_{cc}\rho$ & S   & $4488$ & $\Sigma_c D$ & $4384$ \\
$(4396)$         & H   & $4647$ & $(4324)$     & $4714$ \\
                 & S+H & $4488$ &              & $4384$ \\[2ex]
$\Sigma_c D^*$ & S   & $4503$ & & \\
$(4462)$       & H   & $4627$ & & \\
               & S+H & $4503$ & & \\
%     ~~~~~Mixed (singlet)  & 10195 & &  \\
%     ~~~~~Mixed (full)  & 10195 & & \\  \hline\hline
\end{tabular}
\end{ruledtabular}
\end{table}

\begin{table}[!t]
\caption{\label{Gresult7} The lowest eigen-eneries of $\Sigma^*_c D^*$ with $I(J^P)=\frac32(\frac12^-)$, and the rotated angle $\theta=0^\circ$. (unit: MeV) }
\begin{ruledtabular}
\begin{tabular}{lcccc}
Channel   & Color & $M$ & $E_B$ & $M'$ \\[2ex]
$\Sigma^*_c D^*$ & S   & $4548$ & $-3$   & $4524$ \\
$(4527)$         & H   & $4693$ & $+142$ & $4669$ \\
                 & S+H & $4547$ & $-4$   & $4523$ \\
%     ~~~~~Mixed (singlet)  & 10195 & &  \\
%     ~~~~~Mixed (full)  & 10195 & & \\  \hline\hline
\end{tabular}
\end{ruledtabular}
\end{table}

{\bf The $\bm{I(J^P)=\frac32(\frac12^-)}$ channel:} Among all of the possible channels $\Xi_{cc}\pi$, $\Xi^{(*)}_{cc}\rho$ and $\Sigma^{(*)}_c D^{(*)}$ listed in Tables~\ref{Gresult6} and~\ref{Gresult7}, only $\Sigma_c^* D^*(4523)$ is possibly a bound state, in real-range calculation. Its binding energy is $E_B=-3\,\text{MeV}$ when only the singlet-color channel is considered, and $E_B=-4\,\text{MeV}$ if the coupling with hidden-color channel is included. Therefore, the $\Sigma_c^* D^*$ modified mass is $4523\,\text{MeV}$.

Our results from the coupled-channels calculation within the CSM taking into account a range of rotated angle $\theta \in [0^\circ,6^\circ]$ is shown in Fig.~\ref{PP3}. The distribution of $\Xi_{cc}\pi$ states is the same as that seen in $IJ^P=\frac12 \frac12^-$ case; other channels show continuum-state's behaviour. Let us focus on the middle panel of Fig.~\ref{PP3}, from $4.45-4.62\,\text{GeV}$ energy region, where $\Sigma^{(*)}_c D^{(*)}$ and $\Xi^{(*)}_{cc} \rho$ are established. On one hand, it is clear that the effects of coupled-channels lead to a scattering state of $\Sigma^*_c D^*$ whose original modified bound state mass is $4523\,\text{MeV}$ and the corresponding pole ($E=4547\,\text{MeV}$ in real axis of Fig.~\ref{PP3}) descends gradually with a larger values of the rotated angle $\theta$. On the other hand, an unchanged resonance pole with mass $(E)$ and width $(\Gamma)$ of $4491\,\text{MeV}$ and $2.6\,\text{MeV}$, respectively, is circled with green. We identify this state as a baryon-meson molecule of nature $\Sigma_c D$ with a shifted mass of $4431\,\text{MeV}$ due to the difference between our theoretical and the experimental values of the  $\Sigma_c D$ threshold.

The bottom panel of Fig.~\ref{PP3} shows our results in the energy interval of $4.50$ to $4.53\,\text{GeV}$. One can guess that another possible $\Sigma_c D$ resonance state is found, whose mass and width are $4506\,\text{MeV}$ and $2.2\,\text{MeV}$, respectively. By a mass shift with respect to $\Sigma_c D$, according to previous discussion, the obtained resonance state is $\Sigma_c D(4446)$ with a very small width of $\Gamma=2.2\,\text{MeV}$. As one can elucidate from our discussion until now is that the doubly-charmed pentaquark states present similar features than those hidden-charm ones observed experimentally, $P^+_c(4312)$, $P^+_c(4440)$ and $P^+_c(4457)$~\cite{lhcb:2019pc}, which are mainly explained as molecular states of $\Sigma^{(*)}_c \bar{D}^{(*)}$ configurations~\cite{MZL190311560, JH190311872, YS190400587, CWX190401296, ZHG190400851, HH190400221, HM1904.09756, RZ190410285, MIE190411616, XZW190409891, FLW190503636, LM190504113, ZGW190502892, CFR190410021, FKG190311503, JRZ190410711, CJX190400872, Yang:2015bmv}. We expect that, in the near future, the potential molecular candidates in the doubly-charm sector, $\Sigma_c D(4431)$ and $\Sigma_c D(4446)$, being confirmed experimentally.

%%%%%%%%%%

\begin{table}[!t]
\caption{\label{Gresult8} The lowest eigen-eneries of doubly-charm pentaquarks with $I(J^P)=\frac32(\frac32^-)$, and the rotated angle $\theta=0^\circ$. (unit: MeV) }
\begin{ruledtabular}
\begin{tabular}{lcccc}
Channel   & Color & $M$ & Channel & $M$ \\[2ex]
$\Xi_{cc}\rho$ & S   & $4434$ & $\Xi^*_{cc}\pi$ & $3866$ \\
$(4293)$       & H   & $4708$ & $(3760)$        & $4692$ \\
               & S+H & $4434$ &                 & $3866$ \\[2ex]
$\Xi^*_{cc}\rho$ & S   & $4488$ & $\Sigma_c D^*$ & $4503$ \\
$(4396)$         & H   & $4678$ & $(4462)$       & $4719$ \\
                 & S+H & $4488$ &                & $4503$ \\[2ex]
$\Sigma^*_c D$ & S   & $4432$ & & \\
$(4389)$       & H   & $4695$ & & \\
               & S+H & $4432$ & & \\
%     ~~~~~Mixed (singlet)  & 10195 & &  \\
%     ~~~~~Mixed (full)  & 10195 & & \\  \hline\hline
\end{tabular}
\end{ruledtabular}
\end{table}

\begin{table}[!t]
\caption{\label{Gresult9} The lowest eigen-eneries of $\Sigma^*_c D^*$ with $I(J^P)=\frac32(\frac32^-)$, and the rotated angle $\theta=0^\circ$. (unit: MeV) }
\begin{ruledtabular}
\begin{tabular}{lcccc}
Channel   & Color & $M$ & $E_B$ & $M'$ \\[2ex]
$\Sigma^*_c D^*$ & S   & $4551$ & $0$  & $4527$ \\
$(4527)$         & H   & $4667$ & $+116$ & $4643$ \\
                  & S+H & $4548$ & $-3$  & $4524$ \\
%     ~~~~~Mixed (singlet)  & 10195 & &  \\
%     ~~~~~Mixed (full)  & 10195 & & \\  \hline\hline
\end{tabular}
\end{ruledtabular}
\end{table}

{\bf The $\bm{I(J^P)=\frac32(\frac32^-)}$ channel:} Two almost degenerate bound states of $\Sigma^*_c D^*$ are found among the possible channels: $\Xi^{(*)}_{cc} \rho$, $\Xi^*_{cc} \pi$ and $\Sigma^{(*)}_c D^{(*)}$. As listed in Table~\ref{Gresult9}, these two states $\frac32 \frac12^-$ $\Sigma_c^* D^*$ and $\frac32 \frac32^-$ $\Sigma_c^* D^*$ have masses of $4523\,\text{MeV}$ and $4524\,\text{MeV}$, with binding energies close to $-3\,\text{MeV}$.

With a rotational manipulation for the relative motions of five-quark systems in complex plane, the coupled-channels results are shown in Fig.~\ref{PP10}. Again, the lowest and radial excited states of $\Xi^*_{cc} \pi$ are both scattering ones with theoretical a mass of $3866\,\text{MeV}$ and $4305\,\text{MeV}$, respectively.

A possible resonance state of $\Sigma_c D^*$ is found in the bottom panel of Fig.~\ref{PP10} which is an enlarged part involving the energy interval $4.48-4.65\,\text{GeV}$. Clearly, there are three almost overlapped poles inside the green circle which is above the cut lines of $\Sigma_c D^*$, and the corresponding masses and widths can be cluster around $4555\,\text{MeV}$ and $4.0\,\text{MeV}$ respectively. This resonance can be identified as a $\Sigma_c D^*(4514)$ molecular state whose modified mass $E=4514\,\text{MeV}$ is obtained by a mass shift of $\Delta=41\,\text{MeV}$ according to the calculated results of $\Sigma_c D^*(4462)$ channel in Table~\ref{Gresult8}. Finally, as in the $\frac32 \frac12^- \Sigma^*_c D^*(4523)$ case in coupled-channels calculation, the original bound state of $\Sigma^*_c D^*(4524)$ turned into a scattering one with an unstable pole with a theoretical mass of $4548\,\text{MeV}$ in Fig.~\ref{PP10}.

%%%%%%%%%%

\begin{table}[!t]
\caption{\label{Gresult10} The lowest eigen-eneries of doubly-charm pentaquarks with $I(J^P)=\frac32(\frac52^-)$, and the rotated angle $\theta=0^\circ$. (unit: MeV) }
\begin{ruledtabular}
\begin{tabular}{lcccc}
Channel   & Color & $M$ & Channel & $M$ \\[2ex]
$\Xi^*_{cc}\rho$ & S   & $4488$ & $\Sigma^*_c D^*$  & $4551$ \\
$(4396)$         & H   & $4727$ & $(4527)$ & $4706$ \\
                  & S+H & $4488$ & & $4551$ \\
%     ~~~~~Mixed (singlet)  & 10195 & &  \\
%     ~~~~~Mixed (full)  & 10195 & & \\  \hline\hline
\end{tabular}
\end{ruledtabular}
\end{table}

{\bf $\bm{I(J^P)=\frac32(\frac52^-)}$ channel:} Only two baryon-meson channels contribute to this case: $\Xi_{cc}^* \rho$ and $\Sigma_c^* D^*$. Table~\ref{Gresult10} shows that we do not find any bound state in these two configurations. However, in coupled-channels calculation within complex-scaling, a possible $\Xi^*_{cc} \rho$ resonance state with a small decay width is found. In Fig.~\ref{PP11}, an unchanged pole, circled in green, above the threshold lines of $\Xi^*_{cc} \rho$ appears, and its corresponding mass and width are $4553\,\text{MeV}$ and $3.0\,\text{MeV}$, respectively. Therefore, after a mass shift $\Delta=92\,\text{MeV}$ with respect to the experimental value of $\Xi^*_{cc} \rho$ threshold, the obtained resonance mass is $4461\,\text{MeV}$. It would be interesting to explore the possible existence of this high-isospin and -spin $\Xi^*_{cc} \rho$ resonance, although we understand that it is experimentally challenging.

\begin{table}[!t]
\caption{\label{Rsum} Possible bound and resonance states of doubly charmed pentaquarks. The last column lists, in MeV, either the binding energy of the bound-state or the decay width of the resonance.}
\begin{ruledtabular}
\begin{tabular}{lll}
& Quantum state & $E_B$ (MeV) \\[1ex]
Bound states & $\frac{1}{2}\frac{1}{2}^-$ $\Lambda_c D^*(4291)$  & -2 \\
             & $\frac{1}{2}\frac{3}{2}^-$ $\Sigma_c D^*(4461)$   & -1 \\
             & $\frac{3}{2}\frac{1}{2}^-$ $\Sigma^*_c D^*(4523)$ & -4 \\
             & $\frac{3}{2}\frac{3}{2}^-$ $\Sigma^*_c D^*(4524)$ & -3 \\
             & $\frac{1}{2}\frac{3}{2}^-$ $\Xi^*_{cc} \pi(3757)$ & -3 \\[1ex]
& Quantum state & $\Gamma$ (MeV) \\[1ex]
Resonance states & $\frac{1}{2}\frac{1}{2}^-$ $\Sigma_c D(4356)$      & 4.8 \\
                 & $\frac{1}{2}\frac{3}{2}^-$ $\Sigma^*_c D(4449)$    & 8.0\\
                 & $\frac{3}{2}\frac{1}{2}^-$ $\Sigma_c D(4431)$      & 2.6\\
                 & $\frac{3}{2}\frac{1}{2}^-$ $\Sigma_c D(4446)$      & 2.2 \\
                 & $\frac{3}{2}\frac{3}{2}^-$ $\Sigma_c D^*(4514)$    & 4.0 \\
                 & $\frac{3}{2}\frac{5}{2}^-$ $\Xi^*_{cc} \rho(4461)$ & 3.0 \\
\end{tabular}
\end{ruledtabular}
\end{table}

%%%%%%%%%%%%%%%%%%%%%%%%%%%%%%%%%%%%%%%%%%%%%%%%%%%%%%%%%%%%%%%%%%%%%%%%%%%%%%%%

\section{Epilogue}
\label{sec:summary}

The hidden-charm pentaquark signals $P_c(4380)^+$ and $P_c(4450)^+$ were firstly discovered by the LHCb Collaboration in 2015, and then three new pentaquark states $P_c(4312)^+$, $P_c(4440)^+$ and $P_c(4457)^+$ were also announced by the same collaboration with a much more higher statistical significance in 2019. Extensive theoretical investigations have been devoted to explain these possible $\Sigma^{(*)}_c \bar{D}^{(*)}$ molecular states. In Ref.~\cite{Yang:2015bmv}, within a chiral quark model formalism, the $P_c(4380)^+$ was suggested to be a bound state of $\Sigma_c^\ast\bar{D}$ with quantum numbers $IJ^P=\frac12 \frac32^-$. Furthermore, the three newly observed pentaquark states $P_c(4312)^+$, $P_c(4440)^+$ and $P_c(4457)^+$ can also be identified as molecular states of $J^P=\frac12^-$ $\Sigma_c\bar{D}$, $\frac12^-$ $\Sigma_c\bar{D}^*$ and $\frac32^-$ $\Sigma_c\bar{D}^*$, respectively, belonging all of them to the $\frac12$ isospin sector. Accordingly, with this effective phenomenological model, it is natural to expect a subsequent observation of the doubly charmed pentaquark states within a similar energy range ($4.3$ to~$4.5\,\text{MeV}$).

In the present work, we have systematically studied the possibility of having pentaquark bound- and resonance-states in the doubly-charm sector with quantum numbers $J^P=\frac12^-$, $\frac32^-$ and $\frac52^-$, and in the $\frac12$ and $\frac32$ isospin sectors. The chiral quark model used is based on the existence of Goldstone-boson exchange interactions between light quarks that are encoded in a phenomenological potential which already contains the perturbative one-gluon exchange and the nonperturbative linear-screened confining terms. Note that the model parameters have been fitted in the past through hadron, hadron-hadron and multiquark phenomenology. Within the same framework, there is also a successful explanation to the observed hidden-charm pentaquark states and a prediction of their $P_b^+$ partners. Moreover, the five-body bound, scattering and resonance states problems are accurately solved by means of the Gau\ss ian expansion method along with the complex scaling method.

Several possible bound and resonance states are found for doubly-charm pentaquark states within the scanned quantum numbers: $J^P=\frac12^-$, $\frac32^-$, $\frac52^-$ and $I=\frac12$, $\frac32$. These are characterized by the following features: (i) there are bound states of $\frac12 \frac12^-$ $\Lambda_c D^*(4291)$, $\frac12 \frac32^-$ $\Sigma_c D^*(4461)$, $\frac32 \frac12^-$ $\Sigma^*_c D^*(4523)$ and $\frac32 \frac32^-$ $\Sigma^*_c D^*(4524)$, their binding energies are $-2\,\text{MeV}$, $-1\,\text{MeV}$, $-4\,\text{MeV}$ and $-3\,\text{MeV}$, respectively. However, all of them become a scattering state in coupled-channels calculation, (ii) narrow baryon-meson resonance states are obtained in coupled-channels cases, $\frac12 \frac12^-$ $\Sigma_c D(4356)$, $\frac12 \frac32^-$ $\Sigma^*_c D(4449)$, $\frac32 \frac12^-$ $\Sigma_c D(4431)$, $\frac32 \frac12^-$ $\Sigma_c D(4446)$, $\frac32 \frac32^-$ $\Sigma_c D^*(4514)$ and $\frac32 \frac52^-$ $\Xi^*_{cc} \rho(4461)$, their resonance widths are $4.8$, $8.0$, $2.6$, $2.2$, $4.0$ and $3.0\,\text{MeV}$ respectively, (iii) one $\Xi^*_{cc} \pi(3757)$ bound state with binding energy $E_B=-3\,\text{MeV}$ is identified within the coupled-channels calculation of quantum number $IJ^P=\frac12 \frac32^-$. Note here that the former numbers within parentheses are all of the modified masses.

Last but not  least, based on the success of Ref.~\cite{Yang:2015bmv} in explaining $P_c(4380)^+$ and predicting $P_c(4312)^+$, $P_c(4440)^+$ and $P_c(4457)^+$ hidden-charm pentaquark states, the possible bound and resonance states in doubly-charm sector mentioned above are expected to be identified in future high energy physics experiments.


\begin{acknowledgments}
G. Yang would like to thank L. He for his support and informative discussions. Work supported by: China Postdoctoral Science Foundation Grant no. 2019M650617. National Natural Science Foundation of China under grant No. 11535005, 11775118, 11890712 and 11775123; by Spanish Ministerio de Econom\'ia, Industria y Competitividad under contract No. FPA2017-86380-P; and by Junta de Andaluc\'ia under contract No. UHU-1264517 and PY18-5057.
\end{acknowledgments}

%%%%%%%%%%%%%%%%%%%%%%%%%%%%%%%%%%%%%%%%%%%%%%%%%%%%%%%%%%%%%%%%%%%%%%%%%%%%%%%%

\bibliography{OpencharmPentaquarks}

\end{document}